\documentclass[11pt,a4paper]{article}
\pdfoutput=1
\usepackage{jheppub}

\usepackage{multirow, graphicx,amssymb,url,mathrsfs,amsmath}
\usepackage{wrapfig,boxedminipage,setspace,subfigure,epsfig}
\usepackage{amsxtra,amstext,latexsym,dsfont,amsfonts}
\usepackage{color,eucal}
\usepackage[dvipsnames]{xcolor}
\usepackage{float}
\usepackage{slashed,comment}
\usepackage{kotex}
\usepackage{contour}
\usepackage{tikz}
\usetikzlibrary{calc,patterns,angles,quotes}
\usetikzlibrary{decorations.pathreplacing,decorations.markings,snakes}

\usepackage{tabularx, array}

\usepackage{amssymb}
\usepackage{pifont}

\newcolumntype{L}[1]{>{\raggedright\arraybackslash}p{#1}}
\newcolumntype{C}[1]{>{\centering\arraybackslash}p{#1}}
\newcolumntype{R}[1]{>{\raggedleft\arraybackslash}p{#1}}

\renewcommand{\l}{\lambda}

\newcommand{\Tr}{{\rm Tr}\,}

\usepackage{xparse}
\ExplSyntaxOn
\NewDocumentCommand{\HS}{m}
 {
  \seq_set_split:Nnn \l_tmpa_seq { ~ } { #1 }
  \seq_map_inline:Nn \l_tmpa_seq { \contour{green}{##1} ~ } \unskip
 }
\ExplSyntaxOff

\title{Spread and Spectral Complexity in Quantum Spin Chains: from Integrability to Chaos}

\author[a]{Hugo A. Camargo,}
\author[a,b,c,d]{Kyoung-Bum Huh,}
\author[a]{Viktor Jahnke,}
\author[e,f]{Hyun-Sik Jeong,}
\author[a,g]{Keun-Young Kim} 
\author[h]{and Mitsuhiro Nishida}

\emailAdd{hugo.camargo@gist.ac.kr}
\emailAdd{hkabell1689@gmail.com}
\emailAdd{viktorjahnke@gist.ac.kr}
\emailAdd{hyunsik.jeong@csic.es}
\emailAdd{fortoe@gist.ac.kr}
\emailAdd{nishida124@postech.ac.kr}

\preprint{IFT-UAM/CSIC-24-65}

\affiliation[a]{Department of Physics and Photon Science, Gwangju Institute of Science and Technology,\\123 Cheomdan-gwagiro, Gwangju 61005, Korea}
\affiliation[b]{School of Physics and Astronomy, Shanghai Jiao Tong University, Shanghai 200240, China}
\affiliation[c]{Wilczek Quantum Center, School of Physics and Astronomy, Shanghai Jiao Tong University, Shang-
hai 200240, China}
\affiliation[d]{Shanghai Research Center for Quantum Sciences, Shanghai 201315, China}
\affiliation[e]{Instituto de F\'isica Te\'orica UAM/CSIC, Calle Nicol\'as Cabrera 13-15, 28049 Madrid, Spain}
\affiliation[f]{Departamento de F\'isica Te\'orica, Universidad Aut{\'o}noma de Madrid, 28049 Madrid, Spain}
\affiliation[g]{Research Center for Photon Science Technology, Gwangju Institute of Science and Technology, 123 Cheomdan-gwagiro, Gwangju 61005, Korea}
\affiliation[h]{Department of Physics, Pohang University of Science and Technology, Pohang 37673, Korea}

\abstract{
We explore spread and spectral complexity in quantum systems that exhibit a transition from integrability to chaos, namely the mixed-field Ising model and the next-to-nearest-neighbor deformation of the Heisenberg XXZ spin chain. We corroborate the observation that the presence of a peak in spread complexity before its saturation, is a characteristic feature in chaotic systems. We find that, in general, the saturation value of spread complexity post-peak depends not only on the spectral statistics of the Hamiltonian, but also on the specific state. However, there appears to be a maximal universal bound determined by the symmetries and dimension of the Hamiltonian, which is realized by the thermofield double state (TFD) at infinite temperature. We also find that the time scales at which the spread complexity and spectral form factor change their behaviour agree with each other and are independent of the chaotic properties of the systems. In the case of spectral complexity, we identify that the key factor determining its saturation value and timescale in chaotic systems is given by minimum energy difference in the theory's spectrum. This explains observations made in the literature regarding its earlier saturation in chaotic systems compared to their integrable counterparts. We conclude by discussing the properties of the TFD which, we conjecture, make it suitable for probing signatures of chaos in quantum many-body systems.}

\begin{document}
\maketitle

\section{Introduction}\label{sec:intro}

Over the past twenty years, quantum information theory has acquired a pivotal role in the study of Quantum Field Theory (QFT) and quantum gravity. The Anti-de Sitter (AdS)/ Conformal Field Theory (CFT) correspondence has contextualized this relation within a holographic framework, where certain geometric properties of the bulk spacetime are intricately linked to the entanglement characteristics of the boundary field theory. Among the extensively studied concepts, one that was introduced to go beyond usual spacetime probes and explore the black hole interior is \emph{complexity}.

The role of certain extremal codimension-$0$ and codimension-$1$ surfaces in probing holographic spacetimes beyond the black hole horizon was highlighted almost a decade ago ~\cite{Harlow:2013tf,Stanford:2014jda,Susskind:2014rva,Brown:2015bva,Brown:2015lvg,Czech:2017ryf} and since then, the notion of holographic complexity has been generalized for infinite classes of codimension-$0$ and codimension-$1$ gravitational objects ~\cite{Belin:2021bga,Belin:2022xmt,Jorstad:2023kmq,Myers:2024vve}. All these quantities have two defining features: a continued time-evolution far beyond usual thermalization timescales (usually of the order of the so-called scrambling time $t_\text{scrambling}$) and a universal response to shockwave perturbations that results in a time delay called the switchback effect~\cite{Stanford:2014jda}. The late-time ($t\gg t_\text{scrambling}$) growth of these quantities is furthermore expected to be linear in time.

These properties were inspired by quantum (circuit) complexity, a measure used in quantum computing to quantify the difficulty of preparing quantum states using limited resources (e.g. using only certain allowed operations belonging to a set of universal gates)~\cite{Watrous:2008aa,Aaronson:2016vto,Gharibian:2014aa,Osborne:2011aa}. The central idea being that the quantum complexity of boundary quantum states is computed in the bulk spacetime via such gravitational objects. However, two key issues arise in this identification. The first one is that the late-time linear growth of quantum complexity is expected to transition to a plateau for very late times~\cite{Susskind:2015toa,Brown:2016wib,Balasubramanian:2019wgd,Balasubramanian:2021mxo,Haferkamp:2021uxo}. It remained an open problem to replicate this feature in avatars of holographic complexity, until authors in~\cite{Iliesiu:2021ari} showed that this issue could be resolved in a large class of two-dimensional gravity models (which include Jackiw--Teitelboim (JT) gravity~\cite{Teitelboim:1983ux,Jackiw:1984je,Maldacena:2016upp}) by taking into account non-perturbative contributions to the (renormalized) length of the Einstein--Rosen Bridge (ERB).\footnote{See~\cite{Balasubramanian:2022gmo} for an alternative approach.} A byproduct of this study was the introduction of a new spectral quantity, dubbed \emph{spectral complexity}, which measures the quantum complexity of the state holographically dual to the black hole, the Thermofield Double (TFD) state, and which exhibits this late time saturation.

Perhaps the more pressing issue, however, has been the absence of a precise understanding of quantum complexity in QFTs and establishing a correspondence between ambiguities in the generalized holographic complexity proposals and their potential QFT counterparts. Despite this, there has been important progress in the study of circuit complexity in free and weakly-coupled QFTs~\cite{Jefferson:2017sdb,Chapman:2017rqy,Khan:2018rzm,Hackl:2018ptj,Bhattacharyya:2018bbv} as well as CFTs~\cite{Caputa:2017urj,Caputa:2017yrh,Caputa:2018kdj,Erdmenger:2020sup,Flory:2020dja,Flory:2020eot,Chagnet:2021uvi,Camargo:2022wkd}, using different frameworks that rely on geometric aspects, as well as group-theoretic approaches and the optimization of path-integrals. See~\cite{Chapman:2021jbh} for a review.

Remarkably, a one-to-one correspondence between a specific notion of quantum complexity and the length of the ERB was recently found in~\cite{Rabinovici:2023yex} within the context JT gravity and the double-scaled Sachdev--Ye--Kitaev (SYK) model~\cite{SachdevYeModel,KitaevTalks}. The notion computed in the SYK model is Krylov state complexity, also known as \emph{spread complexity}~\cite{Balasubramanian:2022tpr}. Krylov complexity (or simply K-complexity) was first was introduced in~\cite{Parker:2018yvk} as a measure of operator growth for operators undergoing Heisenberg time evolution in quantum many-body systems with local interactions. It was then extended to study the Schr\"{o}dinger time evolution of states and measure their ``spread'' in a subspace of Hilbert space known as Kryolv space, a space spanned by successive applications of the quantum system's Hamiltonian on an initial state. Spread and K-complexity have been studied in different contexts such as topological and quantum phase transitions~\cite{Caputa:2022eye,Afrasiar:2022efk,Caputa:2022yju,Pal:2023yik}, operator growth~\cite{Barbon:2019wsy,Bhattacharjee:2022vlt}, CFTs~\cite{Dymarsky:2019elm,Dymarsky:2021bjq,Kundu:2023hbk, Malvimat:2024vhr}, open quantum systems~\cite{Bhattacharya:2022gbz,Bhattacharjee:2022lzy,Mohan:2023btr,Bhattacharya:2023zqt,Bhattacharjee:2023uwx,Carolan:2024wov}, and other related contexts~\cite{Yates:2021asz,Caputa:2021ori,Patramanis:2021lkx,Trigueros:2021rwj,Rabinovici:2020ryf,Rabinovici:2021qqt,Rabinovici:2022beu,Bhattacharya:2023xjx,Bhattacharjee:2022qjw,Chattopadhyay:2023fob,Bhattacharjee:2023dik,Bhattacharjee:2022ave,Takahashi:2023nkt,Camargo:2022rnt,Avdoshkin:2022xuw,Erdmenger:2023wjg,Hashimoto:2023swv,Camargo:2023eev,Iizuka:2023pov,Caputa:2023vyr,Fan:2023ohh,Vasli:2023syq,Gautam:2023bcm,Iizuka:2023fba,Huh:2023jxt,Anegawa:2024wov,Caputa:2024vrn,Chen:2024imd,Caputa:2024xkp,Chattopadhyay:2024pdj, Nandy:2023brt,Aguilar-Gutierrez:2023nyk,Aguilar-Gutierrez:2024nau}.
For a comprehensive and recent review of both spread and K-complexity, together with an extensive list of references, we refer the reader to \cite{Nandy:2024htc}.

In both cases, the key aspect underlying the determination of Krylov (operator or state) complexity involves the construction of the Krylov basis, an orthonormal basis in Krylov space.\footnote{In the case of operator growth, the Krylov space is a subspace of the Gelfand--Naimark--Segal (GNS) Hilbert space, constructed after appropriately choosing an inner product of operators. In this context, the GNS Hamiltonian is often called the Liouvillian, a superoperator that acts by commuting operators with the Hamiltonian.} This procedure can be accomplished by using the recursion method~\cite{Lanczos:1950zz,RecursionBook}, known in this context as the Lanczos algorithm. This procedure furthermore yields two sets of coefficients $\lbrace a_{n},b_{n} \rbrace$, known as Lanczos coefficients, which are nothing more than the tri-diagonal elements of the Liouvillian (or Hamiltonian) expressed in the Krylov basis and contain all the information about the dynamics of the operator or state.

Spread and K-complexity have also been at the center of recent explorations of quantum chaos. Of particular interest for this paper, are studies that relate spread complexity with other probes of quantum chaos such as the Spectral Form Factor (SFF)~\cite{Guhr:1997ve,Brezin:1997rze,Cotler:2016fpe} and spectral complexity (see e.g.~\cite{Camargo:2023eev,Erdmenger:2023wjg} for examples in this direction). In~\cite{Erdmenger:2023wjg} authors studied connections between these quantities in the context of random matrix ensembles and {elucidated their interconnections}. The authors also derived an Ehrenfest theorem in Krylov space linking the early-time behavior of the spread complexity for the TFD state with its spectral complexity. Moreover, building upon observations made in~\cite{Balasubramanian:2022tpr} for random matrix ensembles and the SYK model, they provided further evidence that the presence of a late-time peak in spread complexity is a characteristic feature of chaotic systems. 

{Furthermore, it has been reported \cite{Erdmenger:2023wjg} that the time evolution of spread complexity demonstrates a pattern of quadratic growth followed by linear growth, leading to a peak and eventual saturation, whose transition time scales are close to those in the SFF.\footnote{In \cite{Erdmenger:2023wjg}, the initial growth region are referred to as the early-time behavior, while the peak and saturation as the late-time behavior. It is also worth noting \cite{Huh:2023jxt} that the appearance of the peak in the saddle-dominated scrambling suggests that correctly identifying genuine quantum chaos often requires further physical insights in addition to spread complexity.} The authors also suggested that the pronounced ramp in the SFF could be responsible for the emergence of a peak in spread complexity.}

Although spread and spectral complexity were originally proposed in high-energy theoretical physics, these complexities can also be studied in traditional quantum spin chain models in condensed-matter systems. Even without taking disorder average over coupling constants, several spin models exhibit chaotic behavior such as the Wigner--Dyson distribution of level spacing in random matrix theories (RMTs). Since the spectral density of these spin models differs from that of RMTs, the complexities associated with chaotic spin models may offer further insights into quantum chaos, similar to observations in the SYK model.

Motivated by these developments, the goal of our work is to further investigate the role of spread complexity and spectral complexity as probes of quantum chaos in quantum many-body systems. Furthermore, we are interested in understanding the characteristic timescales at which these quantities exhibit their paradigmatic behaviour in chaotic systems, i.e. the peak in spread complexity and the saturation in spectral complexity. In particular, to further elucidate their role in probing quantum chaos, it is imperative to contrast their properties in chaotic and integrable models. To this end, we consider two spin chain models that exhibit a well-studied transition between integrability and chaos: the mixed-field Ising model and a next-to-nearest-neighbour (NNN) deformation of the Heisenberg XXZ model.

It is worth noting that spin chain models serve as important exemplars of quantum mechanical systems, providing a simplified yet insightful depiction of essential quantum phenomena. Through our investigation, we aspire to offer valuable insights into foundational aspects of quantum complexities, contributing to a deeper understanding of many-body quantum physics and facilitating potential advancements in the study of holographic complexity.

\paragraph{Summary of Results and Outline.}
We show that both spread and spectral complexity exhibit distinguishing features in chaotic spin systems which can be directly contrasted with their integrable counterparts. That is: the presence of a peak in spread complexity and a saturation of spectral complexity which is governed by the minimum value of the energy difference in the spectrum. These results can be seen as an effort in the same direction as studies performed for Krylov operator complexity~\cite{Rabinovici:2020ryf,Rabinovici:2021qqt,Rabinovici:2022beu,Bhattacharya:2023xjx}. 
In particular, for spread complexity, we verify that the characteristic peak is only manifest in the chaotic case, and that its late-time behaviour is model dependent. For the chaotic point of the Ising model, we find that the spread complexity saturates to a constant whose value is comparable with the one for the integrable case. In contrast, for the NNN deformation of the Heisenberg chain, we find that the saturation value of the integrable case lies below the value of the chaotic one. This is in direct analogy with the behaviour of Krylov (operator) complexity in the same models.
Finally, for spectral complexity we confirm that its saturation value for the chaotic cases occurs much earlier than in their integrable counterparts. We also determine that this saturation occurs at a characteristic timescale in chaotic systems which is proportional to the inverse of the minimum energy difference in the Hamiltonian's spectrum. This feature is explored for Hamiltonians belonging to different random matrix ensembles. 

Our paper is organized as follows. In Sec.~\ref{sec:prelim} we provide an overview of spread complexity, the SFF and spectral complexity. In Sec.~\ref{subsec:Models} we introduce the spin chain models. Then, in Sec.~\ref{subsec:ResultsSpreadC} we discuss the results for spread complexity and we compare them with the SFF. In Sec.~\ref{subsec:ResultsSpectral} we analyze spectral complexity, examining characteristic timescales and making comparisons with those of random matrices. Finally, Sec.~\ref{sec:Discussion} is dedicated to discussing our findings and drawing conclusions.


\section{Preliminaries}\label{sec:prelim}

\subsection{Spread Complexity}\label{subsec:SpreadC}

In this section we briefly review spread complexity ~\cite{Balasubramanian:2022tpr}. Consider a quantum system with a time-independent Hamiltonian $H$. The unitary time-evolution of a pure state $\vert \psi(t) \rangle$ is governed by the Schr$\text{\"{o}}$dinger equation
\begin{align}\label{sch}
    i \partial_t |\psi(t) \rangle = H |\psi(t) \rangle \,.
\end{align}
The solution to this equation, $|\psi(t) \rangle=e^{-i H t}|\psi(0) \rangle$, can be represented as a formal power series of states obtained by a successive application of the Hamiltonian on an initial state at $t=0$: $| \psi (0) \rangle$. In other words
\begin{align}
    | \psi(t) \rangle = \sum_{n=0}^\infty \frac{(-i t)^n}{n!} | \psi_n \rangle \,,
\end{align}
where $|\psi_n \rangle := H^n | \psi (0) \rangle$. These states span Krylov space, a subspace of the full Hilbert space. However, in general they are not orthogonal to each other or even properly normalized. In order to find an orthonormal basis $|K_n\rangle$ in Krylov space, called the Krylov basis, we apply the Lanczos algorithm~\cite{Lanczos:1950zz}, which is essentially the Gram--Schmidt orthogonalization procedure:
\begin{align}\label{Lanczos algorithm}
    | A_{n+1} \rangle = (H - a_n) | K_n \rangle - b_n | K_{n-1} \rangle \,,
\end{align}
where $a_n$ and $b_n$ are called Lanczos coefficients and are given by
\begin{align}\label{forconclu}
    a_n = \langle K_n | H | K_n \rangle\,, \quad b_n = \langle A_n | A_n \rangle ^{1/2} \,.
\end{align}
The initial conditions and normalization of the Krylov basis elements are given by,
\begin{align}
    |K_0\rangle = |\psi(0) \rangle \,, \quad | K_n \rangle = b_{n}^{-1} | A_n \rangle \,, \quad b_0 := 0 \,.
\end{align}
With this, we can rewrite the Lanczos algorithm ~\eqref{Lanczos algorithm} as 
\begin{align}\label{Lanczos algorithm2}
    H|K_n \rangle = a_n | K_n \rangle + b_{n+1} | K_{n+1} \rangle + b_n | K_{n-1} \rangle \,,
\end{align}
which shows that in the Krylov basis, the Hamiltonian acquires a tri-diagonal form
\begin{align}\label{Lanczos algorithm3}
    \langle K_{m}\vert H|K_n \rangle = a_n \delta_{m,n} + b_{n+1} \delta_{m,n+1} + b_n \delta_{m,n-1}\,.
\end{align}
In infinite dimensional Hilbert spaces, the Krylov space is also infinite dimensional. However, if the dimension of the Hilbert space is finite, the dimension of the Krylov space is also finite, {being smaller than or equal to the Hilbert space dimension}, and in general depends on the dynamics of $H$ and choice of initial state $\vert \psi_{0}\rangle$. In this case, the Lanczos algorithm comes to an end whenever we reach a value of $n$ for which $b_n$ is zero.

The time-evolved state $|\psi(t) \rangle$ can be expressed in terms of the Krylov basis as follows
\begin{align}
    |\psi(t) \rangle = \sum_n \psi_n(t) | K_n \rangle\,,
\end{align}
where $\psi_n(t) := \langle K_n | \psi(t) \rangle$ are functions such that
\begin{align}
    \sum_{n} |\psi_n(t)|^2 := \sum_{n} \,p_n(t) = 1\,.
\end{align} 
By using~\eqref{Lanczos algorithm2} the Schr\"{o}dinger equation~\eqref{sch} then reduces to the following equation describing the hopping of a particle in a one-dimensional chain, 
\begin{align}\label{Lanczos sch}
    i\, \partial_t \psi_n(t) = a_n \psi_n(t) + b_{n+1}\psi_{n+1}(t) + b_n \psi_{n-1}(t)\,.
\end{align}
For a given initial state $|\psi(0)\rangle$ with initial condition $\psi_n(0) = \delta_{n0}$, its spread complexity is defined as
\begin{align}\label{spread complexity}
    C(t) := \sum_n n \,p_n(t) = \sum_n n |\psi_n(t)|^2 \,,
\end{align}
which measures the ``spreading'' of the initial state along the one-dimensional chain.
In this paper, we consider TFD state as the initial state $\vert \psi(0)\rangle$
\begin{align}
    |\psi(0) \rangle := \vert \textrm{TFD}\rangle = \frac{1}{\sqrt{Z(\beta)}}\sum_n e^{-\frac{\beta E_n}{2}} | n\rangle\otimes\vert\,n \rangle\,,
\end{align}
where $Z(\beta):=\textrm{tr}(e^{-\beta H})$ is the thermal partition function with inverse temperature $\beta$ and where $\vert n \rangle$ are energy eigenvalues $H\vert n \rangle=E_{n}\vert n \rangle$.\footnote{{In the presence of symmetries, we construct the TFD state separately for each symmetry sector of the system.}} We choose this initial state, since the peak structure of spread complexity was observed for it previously in the context of RMTs~\cite{Balasubramanian:2022tpr,Erdmenger:2023wjg}.

\subsection{Spectral Form Factor}
\label{subsec:SFF}

The spectral form factor (SFF) is a quantity which has been studied in the context of quantum chaos for several decades (see e.g.~\cite{Guhr:1997ve,Brezin:1997rze}) as well as in the AdS/CFT correspondence (see e.g.~\cite{Cotler:2016fpe}). It is defined via the analytically continued thermal partition function $Z(\beta+it):= \textrm{tr}(e^{-(\beta+it)H})$ according to the formula
\begin{equation}
    \label{eq:SFF}
    \textrm{SFF}(t):=\frac{\vert Z(\beta+it)\vert^2}{\vert Z(\beta)\vert^2}=\frac{1}{Z(\beta)^2}\sum_{m,n}e^{-\beta(E_{m}+E_{n})}e^{i(E_{m}-E_{n})t}~,
\end{equation}
where, as in the previous section, $\beta$ is the inverse temperature and $\lbrace E_{n}\rbrace$ are the discrete energy eigenvalues of the quantum theory with Hamiltonian $H$. Note that the SFF can be interpreted as the survival probability for the time evolved TFD state \cite{delCampo:2017bzr}.
See equations \eqref{eqqq1} and \eqref{eqqq2}. This is an important interpretation to understand the relation between the spectral complexity and the SFF. In essence, the spectral complexity is a repackaging of the information extracted from the SFF, which is defined only for the TFD state. However, the spread complexity can be defined for any other state, not only the TFD state.

The prototypical shape of the SFF$(t)$ in random matrix ensembles as well as in the SYK model follows the structure: slope, dip, ramp and plateau, after appropriately taking a disordered-average of the couplings in the Hamiltonian $H$. It is precisely because of this observation that the SFF has been used as a diagnostic for RMT behaviour and consequently as a probe of quantum chaos~\cite{Cotler:2016fpe}. 

To be more precise, the key observation is that random matrix ensembles, as well as systems which thermalize, exhibit a phenomenon called \emph{spectral rigidity}, in which the number of energy levels within an energy interval of a given size
exhibits small fluctuations. Given that the SFF is sensitive to energy density pair correlations, spectral rigidity is manifested in the form of a linear ramp. This feature is generically not expected in integrable systems, where energy eigenvalues (within a single symmetry sector) are completely independent and thus do not repel each other.

\subsection{Spectral Complexity}\label{subsec:SpectralC}

Here we briefly review spectral complexity~\cite{Iliesiu:2021ari}. As mentioned in the introduction, this quantity first appeared in the study of holographic complexity within the framework of the AdS/CFT correspondence. In particular, its introduction was motivated by the expected saturation of quantum circuit complexity in chaotic systems for times that are exponential in the entropy of the system~\cite{Susskind:2020wwe,Brown:2016wib,Balasubramanian:2019wgd,Balasubramanian:2021mxo,Haferkamp:2021uxo}.

Authors in~\cite{Iliesiu:2021ari} studied the non-perturbative length of maximal codimension-$1$ slices (the ERB) in a wide class of two-dimensional models of gravity, by taking into account the contribution of higher-genus topologies. This led them to introduce a spectral quantity, spectral complexity, {aimed at representing} the holographic dual of the volume of the ER bridge in the boundary theory and which yields the value of the computational complexity of the TFD state. To be precise, for quantum systems with discrete energy levels, the spectral complexity of the TFD state is given by
\begin{equation}
    \label{eq:SpectralComp}
    C_{S}(t)=\frac{1}{d \, Z(2\beta)}\sum_{\substack{n,m \\ E_{n}\neq E_{m}}}\left(\frac{\sin\left(t(E_{m}-E_{n})/2\right)}{(E_{m}-E_{n})/2}\right)^{2}e^{-\beta(E_{m}+E_{n})}~,
\end{equation}
where $Z(\beta)$ is the thermal partition function, $\lbrace E_{n}\rbrace$ are the discrete energy eigenvalues and where $d$ is the dimension of the Hilbert space $\mathcal{H}$. This quantity is expected to have a  quadratic behavior with time\footnote{For future reference, we notice that for small values of $t$, the spectral complexity becomes
\begin{equation}
    C_{S} \approx \frac{t^2}{d \, Z(2\beta)}\sum_{\substack{n,m \\ E_{n}\neq E_{m}}}  e^{-\beta(E_{m}+E_{n})} = \frac{Z(\beta)^2-Z(2\beta)}{d Z(2\beta)} t^2 \,,
\end{equation}
which, at $\beta=0$, takes the form $C_{S} \approx \left(1-\frac{1}{d}\right)t^2$. 
} behavior at early times, which then transitions to a behavior that appears to be linear in time, but in fact, has a small non-zero second derivative that oscillates significantly. After that, the spectral complexity is expected to plateau to a constant for later times ($e^{S}$), past usual thermalization and scrambling times, as supported by computations done in the SYK model.

An interesting observation is that this quantity is directly related to the spectral form factor SFF$(t)$ (see Sec.~\ref{subsec:SFF}). By taking two time derivatives of the spectral complexity~\eqref{eq:SpectralComp},  one finds~\cite{Erdmenger:2023wjg}
\begin{equation}\label{eq:SFFandSC}
\frac{\textrm{d}^{2}}{\textrm{d}t^{2}}C_{S}(t)=\frac{2 Z(\beta)^2}{d\, \, Z(2\beta)}\textrm{SFF}(t)-\frac{2}{d} \,,
\end{equation}
where we assume that there is no degeneracy of the energy spectrum. This assumption is always true if we focus on a specific symmetry sector of the Hamiltonian.

This quantity has recently been studied in the context of Krylov complexity and quantum chaos~\cite{Alishahiha:2022anw,Erdmenger:2023wjg,Camargo:2023eev}. In~\cite{Alishahiha:2022anw}, authors studied a relation between spectral complexity and spread complexity in the continuum limit. On the other hand, in~\cite{Erdmenger:2023wjg}, it was noted that the spectral complexity of the TFD state in certain RMTs provides an early-time approximation of the spread complexity of the TFD state, as can be seen by an Ehrenfest theorem in Krylov space. Finally, in~\cite{Camargo:2023eev} authors studied this quantity in the context of certain non-integrable quantum mechanical models (quantum billiards). In particular, authors showed that spectral complexity plateaus to different values at different time scales depending on the integrable properties of the system, with the saturation value of this quantity for a circular (integrable) billiard being orders of magnitude larger than for stadium (non-integrable) billiards.

We end this section with a remark. Even though the equivalence between spread and spectral complexity is expected to be satisfied only at early times (and possibly only for certain states such as the TFD), the latter is an interesting quantity that is expected to be also sensitive to the energy spectrum of the theory in a non-trivial way. In particular, much like SFF, it is sensitive to energy-density pair correlations from energy levels which are arbitrarily separated in the spectrum. To see this, consider the long-time average
\begin{align}\label{spectral_late_time}
\lim_{T\to\infty}\frac{1}{T}\int_0^T \textrm{d}t  \,C_{S}(t)=\frac{2}{d\, Z(2\beta)}\sum_{E_{n}\neq E_{m}}\frac{e^{-\beta(E_{m}+E_{n})}}{(E_{m}-E_{n})^2}.
\end{align}
If $C_{S}(t)$ saturates at late times, the saturation value is given by this long-time average. In particular, at infinite temperature $\beta=0$, the saturation value depends only on the energy difference $E_{m} - E_{n}$, not on $E_{m} + E_{n}$.
As a consequence, features of energy level spacing in non-integrable systems such as spectral rigidity should be captured by the saturation value of $C_{S}(t)$. Furthermore, while there is \emph{a priori} no direct relation between the saturation values of spectral and spread complexity for chaotic systems at late times, we believe that the difference in saturation values of spectral complexity in integrable and non-integrable models is sufficient to motivate the idea that this quantity may also be used to study the chaotic properties of quantum many-body systems on its own.

\section{Complexity in Spin Systems}\label{sec:Complexities}

\subsection{The Spin Chain Models}\label{subsec:Models}

In this section we briefly review the spin chain models that we will focus on in this work.

\subsubsection{The mixed-field Ising Model}\label{subsubsec:Ising}

The mixed-field Ising model is a one-dimensional spin chain that includes longitudinal $(h_z)$ and transverse $(h_x)$ magnetic fields. The Hamiltonian can be written as
 \begin{align}\label{Ising}
    H = -\sum^{N-1}_{i=1} S^{z}_{i}S^{z}_{i+1} - \sum^N_{i=1}\left( h_x S^{x}_{i} + h_z S^{z}_{i} \right) \,,
\end{align}
where $S^{x}_i$ and $S^{z}_i$ spin-$1/2$ operators at site $i$, defined as:
\begin{align}
    \label{spinops}
    S^{k}_{i}=\left(\mathds{1}_{2}\right)^{\otimes(i-1)}\otimes \sigma_{k}\otimes\left(\mathds{1}_{2}\right)^{\otimes(N-i)}~,
\end{align}
for $k\in\lbrace x,y,z\rbrace$, and where $\sigma_{k}$ represents the usual Pauli matrices. We consider open boundary conditions, and for a fixed number of sites $N$, the dimension of the Hilbert space is $d=2^N$. 

It is well-known (see e.g.~\cite{Ba_uls_2011, Craps:2019rbj}) that this model exhibits chaotic and integrable properties, according to their spectral statistics, when the longitudinal and transverse fields take the values as\footnote{The mixed-field Ising model also becomes integrable if $h_x=0$.}
%
\begin{align}
(h_x,\,h_z) = 
\begin{cases}
\,\,  (-1.05,\,0.5) \,, \qquad\, (\text{Chaotic case})  \\
\,\, (-1,\,0) \,. \qquad\qquad\,\, (\text{Integrable case}) 
\end{cases}
\end{align}
For any values of $(h_x,h_z)$, this model has parity symmetry -- the Hamiltonian commutes with the parity operator
\begin{equation} \label{eq:parity}
    \hat{\Pi}=
    \begin{cases}
       \hat{P}_{1,N} \, \hat{P}_{2,N-1} \, \cdots \, \hat{P}_{\frac{N}{2},\frac{N+2}{2}} \,,\,\,\quad \text{for}\,\, N \,\, \text{even}\\
       \hat{P}_{1,N} \, \hat{P}_{2,N-1} \, \cdots \, \hat{P}_{\frac{N-1}{2},\frac{N+3}{2}} \,,\,\,\, \text{for}\,\, N \,\, \text{odd}
    \end{cases}
\end{equation}
where the permutation operator
\begin{equation}
    \hat{P}_{i,j}=\frac{1}{2}\left(\mathds{1}_{d}+S_i^xS_j^x+S_i^yS_j^y+S_i^zS_j^z \right)
\end{equation}
permutes the spin configuration of the sites $i^\text{th}$ and $j^\text{th}$. The action of the parity operator on a given state can be understood by imagining a mirror at one edge of the chain. The action of $\hat{\Pi}$ on a given spin configuration (state) returns the mirror image of that spin configuration.
For instance, $\hat{\Pi} \,\,\vert\!\!\uparrow \uparrow \uparrow \downarrow \rangle =\vert\!\!\downarrow \uparrow \uparrow \uparrow  \rangle$. Note that from these two states we can build states with definite parity: $\vert \psi_{\pm}\rangle = \frac{1}{\sqrt{2}}\left( \vert \!\uparrow \uparrow \uparrow \downarrow \rangle \pm \vert\! \downarrow \uparrow \uparrow \uparrow  \rangle \right)$, which are eigenstates of the parity operator $\hat{\Pi} \,\,\vert \psi_{\pm}\rangle=\pm \vert \psi_{\pm}\rangle$ with eigenvalue $\pm 1$.

\begin{figure}[]
 \centering
     {\includegraphics[width=6.6cm]{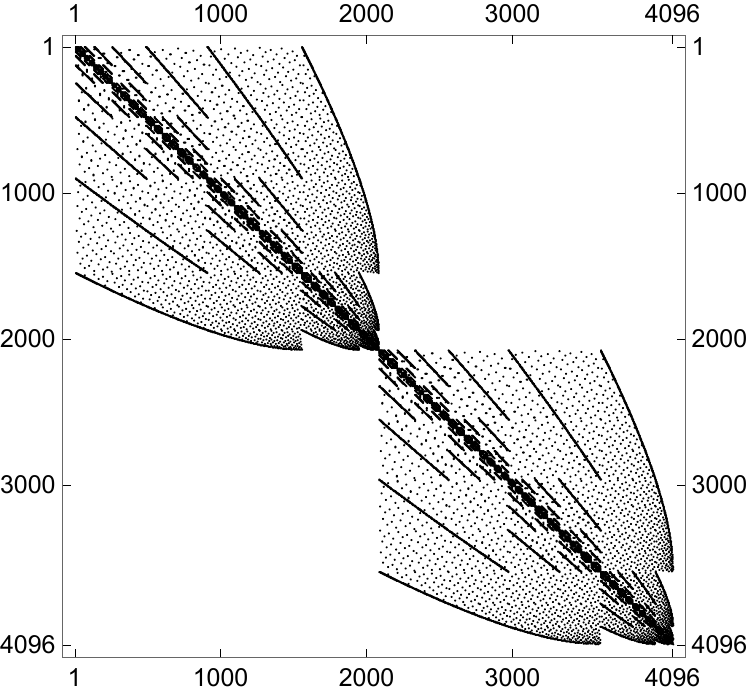} }
\qquad
     {\includegraphics[width=6cm]{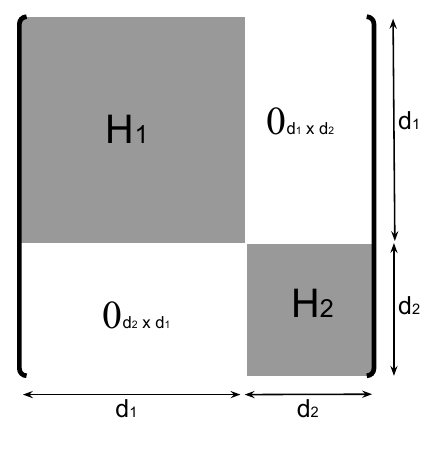} }
     \caption{Block-diagonalized Hamiltonian of the mixed-field Ising model~\eqref{Ising} for $N=12$. \textbf{Left}: Matrix plot -- 
 visual representation showing the places where the Hamiltonian takes non-zero values. \textbf{Right}: Schematic picture showing the size of each block. $H_1$ ($H_2$) is the Hamiltonian in the sector of states with positive (negative) parity. Note that $d_1+d_2=d=2^{N}$.} 
     \label{symmetry}
\end{figure}
\begin{figure}[]
 \centering
     \subfigure[$H_1: (h_x,h_z) = (-1.05,0.5)$]
     {\includegraphics[width=6.3cm]{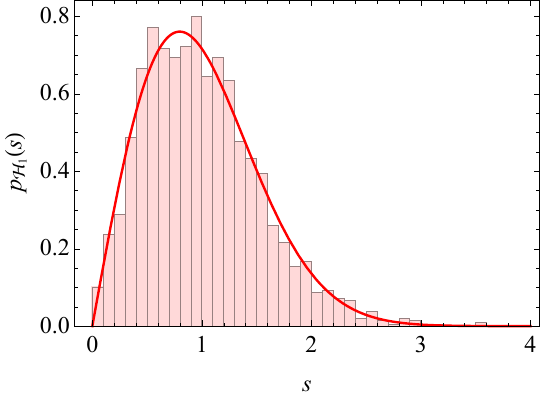} }
     \subfigure[$H_1: (h_x,h_z) = (-1,0.001)$]
     {\includegraphics[width=6.3cm]{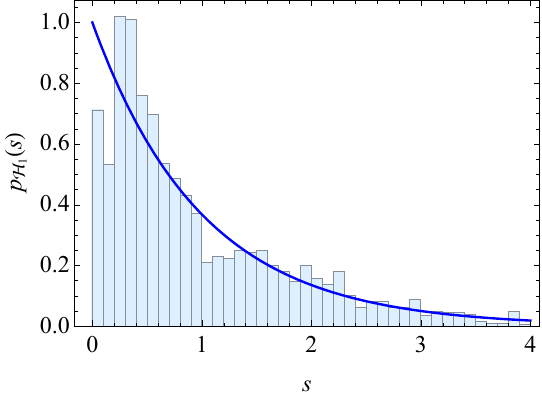} }
         \subfigure[$H_2: (h_x,h_z) = (-1.05,0.5)$]
     {\includegraphics[width=6.3cm]{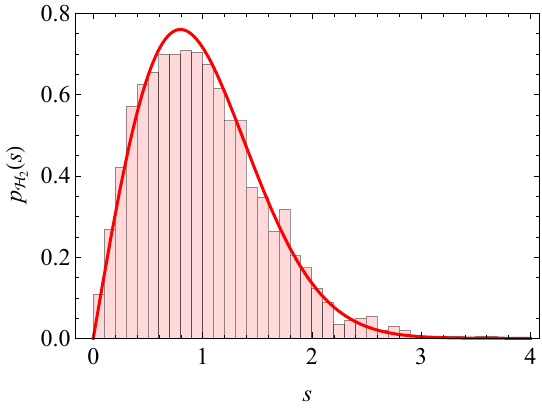} }
     \subfigure[$H_2: (h_x,h_z) = (-1,0.001)$]
     {\includegraphics[width=6.3cm]{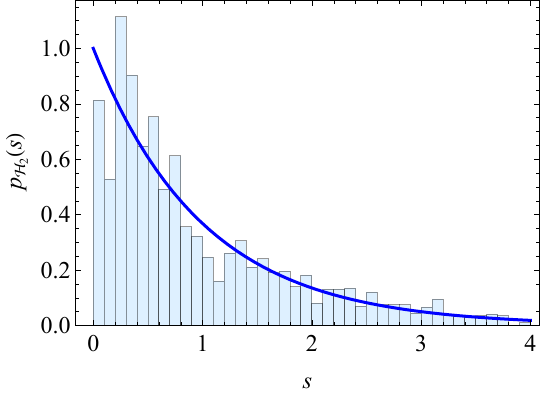} }
     \caption{Nearest-neighbor level spacing distribution in the mixed-field Ising model for the even(odd) symmetry sector $H_1$($H_2$)  with $N=12$ sites. The red line is the Wigner--Dyson distribution (GOE) and the blue line is the Poisson distribution.}
     \label{NNSD}
\end{figure}
Fig.~\ref{symmetry} displays the block-diagonalized Ising Hamiltonian for $N=12$ according to the reflection symmetry around one of the edges of the chain.  Given that there are only two sectors for this symmetry, the Hamiltonian is block-diagonalized into only two blocks, the even and odd parity sectors. These two blocks as labelled as $H_1$ and $H_2$, respectively, as shown in the schematic picture on the right side.

%
%
%

For $h_z=0$, the model develops a new symmetry -- the energy levels do not change if we flip every spin in the chain. This happens because the operator $\Pi_n S_x^{(n)}$ commutes with the Hamiltonian if $h_z=0$~\cite{Craps:2019rbj}. In our numerical calculations for the mixed-field Ising model, we would like to consider both the integrable and the chaotic cases. However, to avoid the emergence of new symmetry sectors at $h_z=0$, we consider a point very close to the integrable point, with $h_z$ very small but nonzero, specifically, we choose $(h_x,h_z)=(-1,0.001)$. 

Two important points that should be taken into account in order to properly recover these spectral properties are the local density of states and the symmetries of the Hamiltonian~\eqref{Ising}. First, one needs to normalize the model-dependent local density of states in order to obtain universal features. This is typically referred to as unfolding of the spectrum. Secondly, if the Hamiltonian has symmetries then one needs to block-diagonalize it according to the conserved charges. This is because only the energy eigenvalues within each individual block are correlated with each other~\cite{DAlessio:2016aa,Craps:2019rbj}.

In Fig.~\ref{NNSD} we display the level spacing distributions for the different symmetry blocks $H_1$, $H_2$ for the integrable and chaotic points of the Hamiltonian. We performed the spectrum unfolding in both cases. In both Figures, the plot on the left represents the nearest-neighbor level spacing statistics for the chaotic case, where the red line is the Wigner--Dyson distribution, while the plot on the right represents the integrable case, where the blue line is the Poisson distribution. As can be seen from these figures, there is a clear distinction between the chaotic and integrable cases for the individual blocks.

\subsubsection{The Heisenberg XXZ Model}\label{subsubsec:Heisenberg}

In this section we briefly discuss the Heisenberg XXZ spin chain and its non-integrable deformation~\cite{Samaj:2013yva}. In integrable systems such as the XXZ spin chain model, it is known that the level spacing distribution follows the Poisson distribution since the energy eigenvalues are uncorrelated. The integrability of such a spin system can be broken by considering the coupling between next-to-nearest-neighbors (NNN)~\cite{Gubin_2012,Rabinovici:2022beu}. The Hamiltonian of this deformation of the XXZ Hamiltonian is given by
\begin{align}\label{XXZ_Ham}
    H = H_{XXZ} + H_{NNN}\,,
\end{align}
where 
\begin{align} \label{eq-hamiltonianXXZ}
    H_{XXZ} &= \sum^{N-1}_{i=1}\,J\, (S^{x}_{i}\,S^{x}_{i+1} + S^{y}_{i}\,S^{y}_{i+1}) + J_{zz}\, S^{z}_i \,S^{z}_{i+1}\,,\cr
    H_{NNN} &=  \sum^{N-2}_{i=1} \, J_c\, S^{z}_{i}\,S^{z}_{i+2}\,,
\end{align}
where $S^{k}_{i}$ are the same spin-$1/2$ operators at site $i$ as in the mixed-field Ising Hamiltonian~\eqref{spinops}.
Note that $J_c$ is the next-to-nearest-neighbors coupling.
The dimension of the Hilbert space in this case is also $2^N$. We set the parameters as follows $(J, \, J_{zz})=(1, \, 0.5)$, and
\begin{align}
(J, \, J_{zz}, \, J_c) = 
\begin{cases}
\,\,  (1, \, 0.5, \, 1) \,, \qquad (\text{Chaotic case}),  \\
\,\,  (1, \, 0.5, \, 0) \,, \qquad (\text{Integrable case}). 
\end{cases}
\end{align}
For any values of $J$, $J_{zz}$ and $J_c$, this model displays parity symmetry, because $[H,\hat{\Pi}]=0$, with $\hat{\Pi}$ given by (\ref{eq:parity}). Additionally, the model conserves the total spin in the $z$-direction, $M_z =\sum_{i=1}^N S_i^z$. For a detailed and pedagogical discussion of the symmetries of this model, we refer the reader to~\cite{Joel2013}.

After fixing $N=15$, and focusing on the $M_z=-5/2$ sector, we checked that the level spacing statistics follows a Wigner--Dyson (Poisson) distribution in the chaotic (integrable) case for both parity sectors. We used the same set of parameters in the analysis below for spread and spectral complexity.

\subsection{Spread Complexity}\label{subsec:ResultsSpreadC}

In this section we study the spread complexity of the TFD state for the mixed-field Ising and the Heisenberg chains. We will consider both the finite-temperature case, as well as the infinite temperature limit $\beta=0$ in which the TFD state becomes the maximally-entangled state
\begin{align}
  \lim_{\beta \rightarrow 0}  \vert \textrm{TFD} \rangle \,\,\longrightarrow\,\, \frac{1}{\sqrt{N}}\sum_{n}\vert n \rangle \otimes \vert n\rangle~.
\end{align}
While we are mostly interested in the spread complexity of the TFD state, we will also analyze the Lanczos coefficients. We first consider the infinite temperature ($\beta = 0$) limit of the TFD state. The finite temperature analysis is shown at the end of this section.

\subsubsection{Lanczos Coefficients and Peak}\label{subsubsec:ResultsSpreadCPeak}

\begin{figure}[]
 \centering
     \subfigure[$H_1: a_n$]
     {\includegraphics[width=6.6cm]{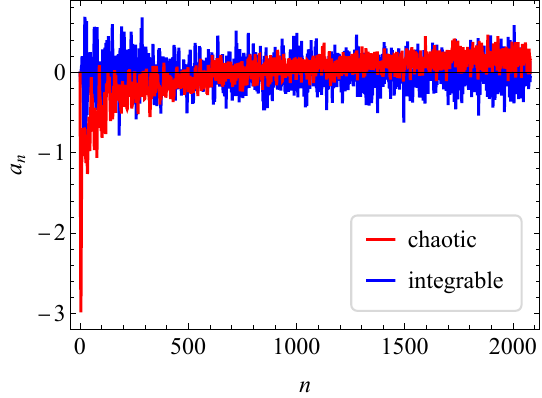} }
     \subfigure[$H_1: b_n$]
     {\includegraphics[width=6.6cm]{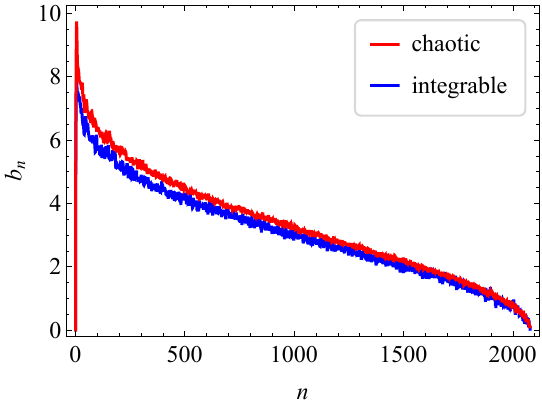} }
     \subfigure[$H_2: a_n$]
     {\includegraphics[width=7.0cm]{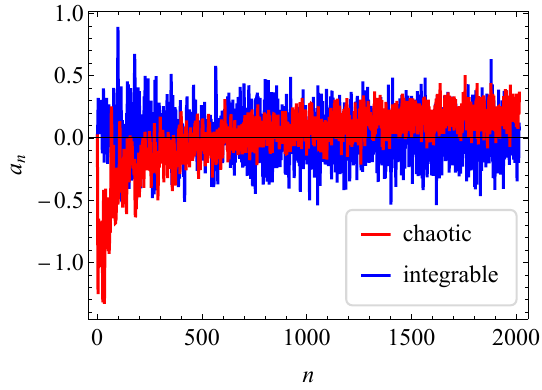} }
     \subfigure[$H_2: b_n$]
     {\includegraphics[width=6.6cm]{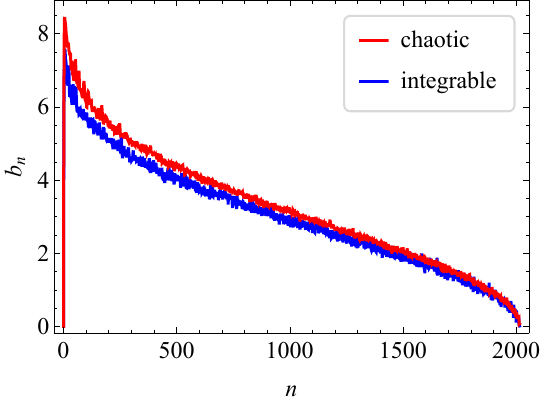} }
     \caption{Lanczos coefficients of the TFD state in the $H_1$ and $H_2$ block of the mixed-field Ising model for $N=12$ and $\beta=0$. The blue data denotes the integrable case, while the red data represents the chaotic one.}
     \label{LanczosH1}
\end{figure}

\paragraph{Mixed-field Ising Chain.}
Fig.~\ref{LanczosH1} display the Lanczos coefficients for the mixed-field Ising for different choices of the parameters $(h_x,h_z)$ and for the different symmetry blocks $H_1$ and $H_2$. 
%
%
%
The blue line represents the almost integrable case for $(h_x,h_z) = (-1,0.001)$, while the red line represents the chaotic one for $(h_x,h_z) = (-1.05, 0.5)$. {The results from the chaotic case resemble behaviors observed in the SYK model and RMT behaviour~\cite{Balasubramanian:2022tpr}, with $a_n$ showing a rapid increase followed by saturation, and $b_n$ displaying a rise followed by a gradual decline. Additionally, one can find that the variance of Lanczos coefficients can be greater in the integrable case compared to the chaotic case, which aligns with findings in~\cite{Hashimoto:2023swv}.}

\begin{figure}[]
 \centering
     \subfigure[$H_1$]
     {\includegraphics[width=6cm]{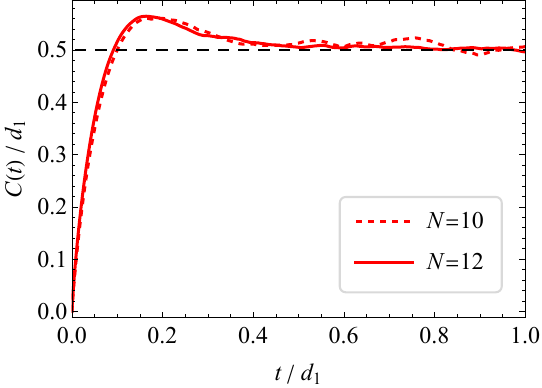} }
     \subfigure[$H_2$]
     {\includegraphics[width=6cm]{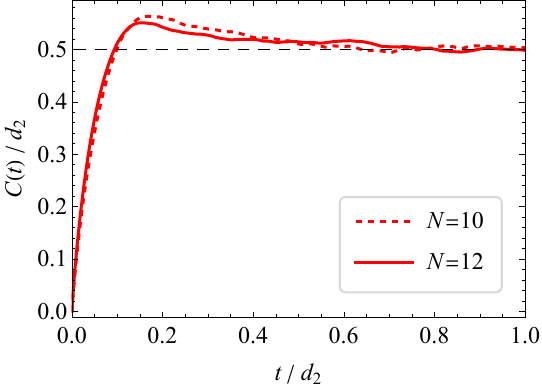} }
     \caption{Spread complexity of the TFD state in the mixed-field Ising model within each symmetry sector for $N=10$ (red dashed line) and $N=12$ (red solid line) when $\beta=0$. Here, $d_i$ is a scaling factor given by the dimension of each symmetry sector $H_i$. This figure shows that if the matrix size $N$ is large enough, then there is no $N$ dependence on the normalized spread complexity. The black dashed line is the asymptotic value \eqref{spread_late_time} of spread complexity at late times. ($d_1=2080\,, d_2=2016$)}
     \label{MFI_spread_dN}
\end{figure}

In Fig.~\ref{MFI_spread_dN}, we display the spread complexity of the TFD state as a function of time for the chaotic case $(h_x,h_z) = (-1.05,0.5)$ for $N=10$ (red dashed line) and $N=12$ (red solid line) when $\beta=0$. 
Note that numerical results have been normalized by the system size $d_i$ ($i=1,2$), and thus the behaviour of the normalized spread complexity does not depend on the number of sites $N$. This aligns with the findings of ~\cite{Balasubramanian:2022tpr}, which reported that for sufficiently large systems, the normalized spread complexity does not depend much on the system size (in our case $N$) when plotted against $t/d_i$, being universal in that sense. In our computation, a spin chain with $N=10$ is large enough to exhibit such universal behavior.
Here we also verify that the late-time behavior (black dashed line) of spread complexity can be estimated by the system size of Hamiltonian ($d_i$): 
\begin{align}\label{spread_late_time}
    C(t=\infty) \approx \frac{d_i - 1}{2}\,,
\end{align}
which is derived for a maximally-entangled state such as the TFD state with $\beta = 0$~\cite{Erdmenger:2023wjg}.


\begin{figure}[]
 \centering
     \subfigure[$H_1$]
     {\includegraphics[width=6cm]{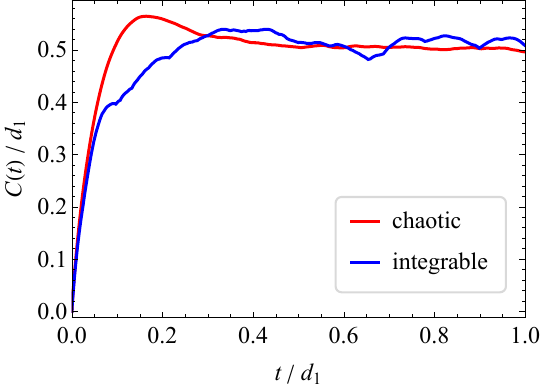} }
     \subfigure[$H_2$]
     {\includegraphics[width=6cm]{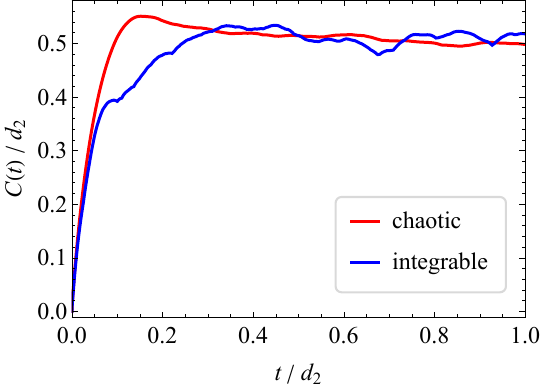} }
     \caption{Spread complexity in the mixed-field Ising model within each symmetry block for $N=12$. The red line is the chaotic case and the blue line is the integrable case.}
     \label{MFI_spread_N12}
\end{figure}
\begin{figure}[]
 \centering
    \subfigure[Chaotic case]
     {\includegraphics[width=6cm]{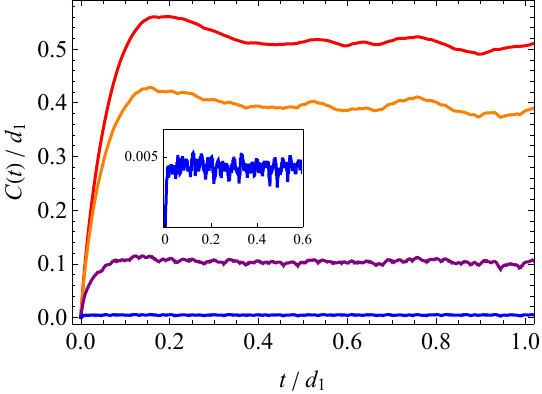} }
     \subfigure[Integrable case]
     {\includegraphics[width=6cm]{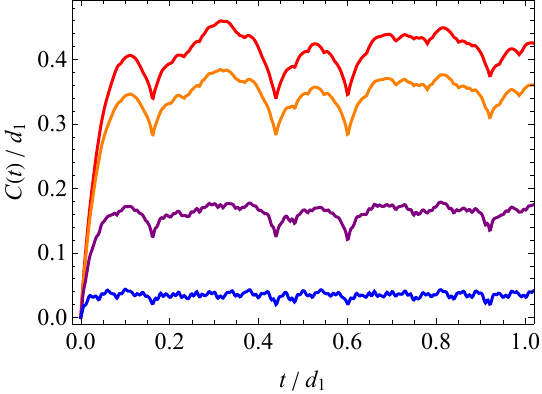} }
     \caption{Temperature dependence of spread complexity with finite $\beta = 0,\,0.2,\,0.5,\,1$ (red, orange, purple, blue) when $N=10$ for the mixed-field Ising.}
     \label{spread_beta_dependence}
\end{figure}

In Fig.~\ref{MFI_spread_N12}, we present the difference in time evolution of spread complexity between the chaotic (red line) and integrable (blue line) cases. This figure shows that the spread complexity presents a characteristic peak for the chaotic case (red line), while it is absent for the integrable case (blue line).  It should be emphasized that taking the symmetry block $H_i$ is crucial for the appearance of the characteristic peak of spread complexity in the chaotic case. In Appendix \ref{AppA}, we show that the peak is less evident if we consider all energy eigenvalues without choosing a symmetry sector.

In addition, Fig.~\ref{MFI_spread_N12} shows that the behaviour of the normalized spread complexity in the mixed-field Ising model is qualitatively the same regardless of the choice of the symmetry sector. For this reason, we focus on one symmetry block $H_1$ in the following discussions.

Finally, we discuss the temperature dependence in the spread complexity of the mixed-field Ising model.
In Fig.~\ref{spread_beta_dependence}, we find that finite temperature ($\beta\neq0$) suppresses spread complexity in both chaotic and integrable cases, and it also diminishes the visibility of the peak in the chaotic case. These findings align with those discussed in RMT and SYK models~\cite{Balasubramanian:2022tpr}. For further explanation on the finite temperature effect, see the discussion on the specific role of TFD state in Sec. \ref{sec:Discussion}. In short, for finite $\beta$, not every energy eigenstate equally contributes to the dynamics of the spread complexity, which then cannot explore the full chaotic features of the spectrum of the Hamiltonian.

\paragraph{NNN Deformation of the Heisenberg Chain.}
The results for the NNN deformation of the Heisenberg chain are not qualitatively different from those of the mixed-field Ising model, which provides further evidence that these quantities are useful to probe signatures of chaos and potentially distinguish chaotic and integrable quantum many-body systems. Because of this, we summarize only the key results
for $\beta=0$.

\begin{figure}[]
 \centering
     \subfigure[$\mathcal{H}_1$]
     {\includegraphics[width=6.6cm]{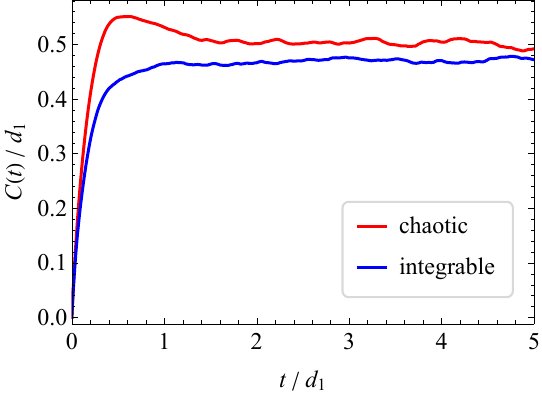} }
     \subfigure[$\mathcal{H}_2$]
     {\includegraphics[width=6.6cm]{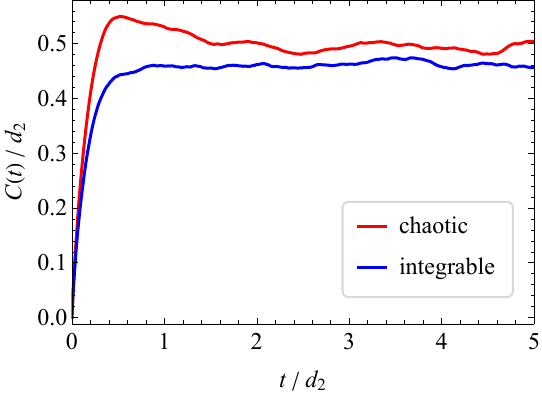} }
     \caption{Spread complexity in NNN deformation of the XXZ spin chain within each symmetry block for $N=15$ and total spin in the $z$-direction $M_z=-5/2$. The red line represents the chaotic case, while the blue line represents the integrable case. }
     \label{XXZ_spread}
\end{figure}

As mentioned previously, the NNN deformation of the Heisenberg Hamiltonian~\eqref{XXZ_Ham} also has two parity symmetry sectors, similarly to the mixed-field Ising model. The results for the spread complexity are shown in Fig.~\ref{XXZ_spread}. Here, as shown in Fig.~\ref{MFI_spread_N12}, the peak structure is also present in the chaotic case. At the same time, the late-time behaviour is also captured by~\eqref{spread_late_time} for the chaotic case.\footnote{\label{footnote7ref}
Nevertheless, it is not surprising that \eqref{spread_late_time} fails to capture the late time behavior of the integrable case (blue data) depicted in Fig. \ref{XXZ_spread}. It has already been demonstrated \cite{Huh:2023jxt} that the integrable models, such as LMG models, cannot adhere to \eqref{spread_late_time}. In the derivation of \eqref{spread_late_time}, it is assumed that there is no energy degeneracy. In numerical calculations on integrable systems, approximate energy degeneracy within numerical precision may occur due to the Poisson distribution of the level spacing. For this reason, within the numerical precision and time scale of our numerical calculations, \eqref{spread_late_time} in the integrable case does not seem to hold in Fig.~\ref{XXZ_spread}.
} Note that the saturation value for the integrable case is noticeably smaller that the chaotic case, in contrast with the Ising chain (cfr. Fig.~\ref{MFI_spread_N12})~\cite{Balasubramanian:2022tpr,Erdmenger:2023wjg}.

Considering that spread complexity quantifies the extent to which the initial state propagates across the Krylov basis, it is somewhat intuitive that the saturation value of spread complexity is greater in chaotic scenarios compared to integrable ones. This implies that a larger portion of the Krylov space is utilized in chaotic cases compared to integrable ones. This aspect could potentially serve as an advantageous characteristic of spread complexity.

\subsubsection{A Comparison with the Spectral Form Factor}\label{subsubsec:ResultsSpreadCvsSFF}

There is an interesting conjecture about similarities between the transition time-scales of spread complexity and the SFF. As discussed in Sec.~\ref{subsec:SFF} the qualitative structure of the SFF is: slope, dip, ramp and plateau, while spread complexity is conjectured to follow a similar pattern: quadratic growth, linear growth, peak and plateau for the RMT or SYK models~\cite{Balasubramanian:2022tpr,Erdmenger:2023wjg}. The goal of this section is to discuss this finding in our spin chain models.

\begin{figure}[]
 \centering
     \subfigure[Chaotic case]
     {\includegraphics[width=6.6cm]{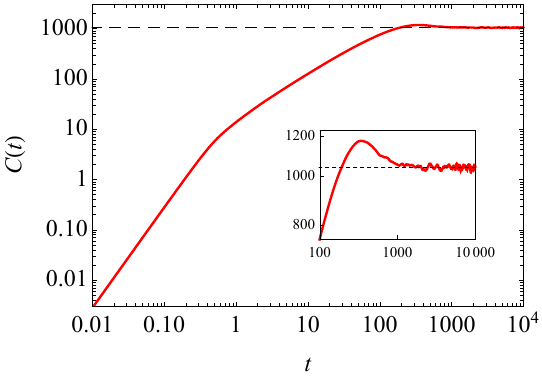} }
     \subfigure[Integrable case]
     {\includegraphics[width=6.6cm]{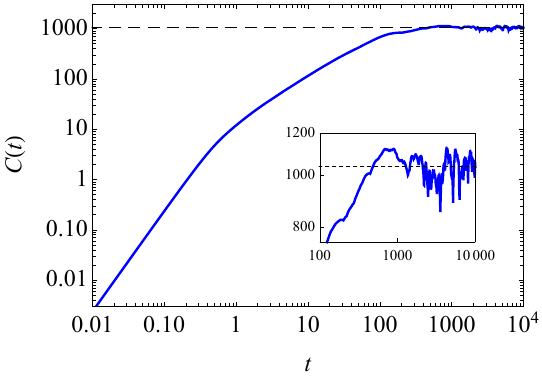} }
     \caption{Log-scale plot of spread complexity in the mixed-field Ising model with $\beta=0$ and $N=12$ for the positive parity sector $H_1$.  The dashed lines are \eqref{spread_late_time} ($d_1=2080\,, d_2=2016$). In the left panel, the peak can be seen in the inset.}
     \label{MFI_spread_Log}
\end{figure}

\begin{figure}[]
 \centering
     \subfigure[Chaotic case]
     {\includegraphics[width=6.6cm]{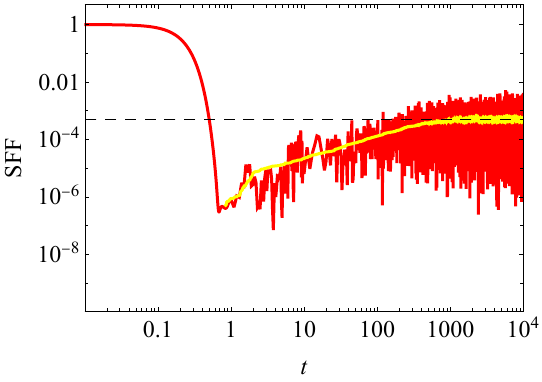} }
     \subfigure[Integrable case]
     {\includegraphics[width=6.6cm]{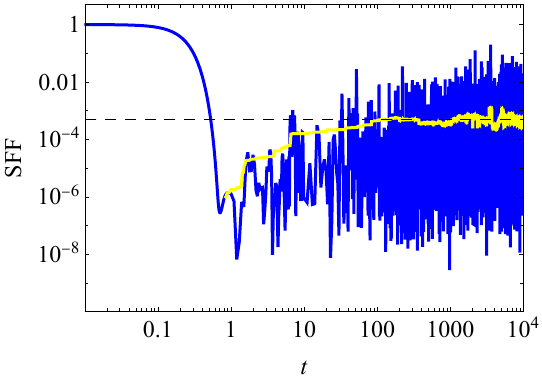} }
     \caption{Log-scale plot of the SFF in the mixed-field Ising model with  $\beta = 0$ and $N=12$ for the positive parity sector $H_1$. The dashed lines are \eqref{MFI_SFF_late_time} ($d_1=2080\,, d_2=2016$). The yellow line represents the moving average, which we used to minimize numerical fluctuations.}
     \label{MFI_SFF}
\end{figure}

\paragraph{Mixed-field Ising model.}
In Figs.~\ref{MFI_spread_Log} and \ref{MFI_SFF} we respectively display the spread complexity and SFF of the mixed-field Ising model.
Our findings align with the argument in \cite{Erdmenger:2023wjg}, suggesting that the transition time scales in both spread complexity and SFF are closely related. For example, we observe that the timescale of the peak in spread complexity and that of the plateau in the SFF both scale linearly with the system's size $d$. See Fig.~\ref{fig:timescale-comparison} for a detailed comparison.

Additionally, a ``clear" ramp in the SFF (yellow line\footnote{The yellow line represents the moving average, which is obtained by averaging over a sliding time window. This procedure smooths out short-term variations and extracts longer-term trends in the data.} in the left panel in Fig. \ref{MFI_SFF}) may result in a peak in spread complexity following its linear growth: see that the ramp in SFF for the integrable case lacks clarity.
\begin{figure}[]
 \centering
     \subfigure[Chaotic case]
     {\includegraphics[width=6.6cm]{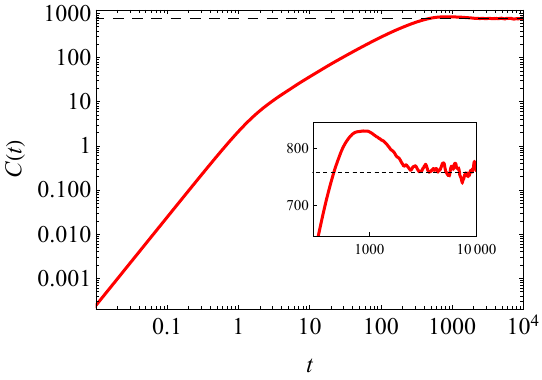} }
     \subfigure[Integrable case]
     {\includegraphics[width=6.6cm]{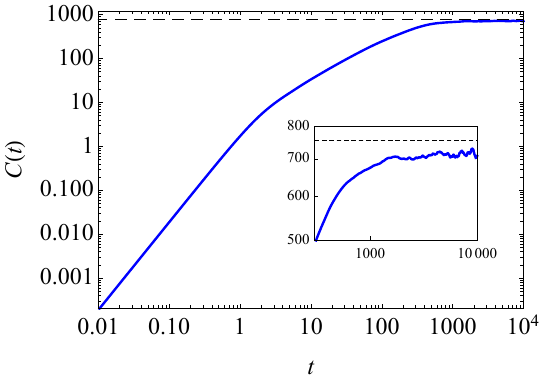} }
     \caption{
     Log-scale plot of spread complexity in the XXZ spin chain with $\beta=0$ and $N=15$ for the positive parity sector $H_1$ with total spin in the $z$-direction $M_z=-5/2$. The dashed lines are \eqref{spread_late_time} ($d_1=1512\,, d_2=1491$). In the left panel, the peak can be seen in the inset.}
     \label{XXZ_Log_spread}
\end{figure}
\begin{figure}[]
 \centering
     \subfigure[Chaotic case]
     {\includegraphics[width=6.6cm]{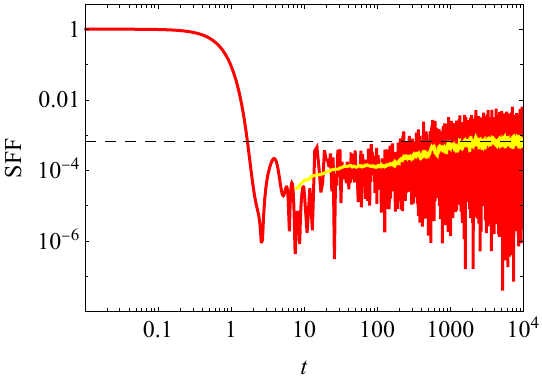} }
     \subfigure[Integrable case]
     {\includegraphics[width=6.6cm]{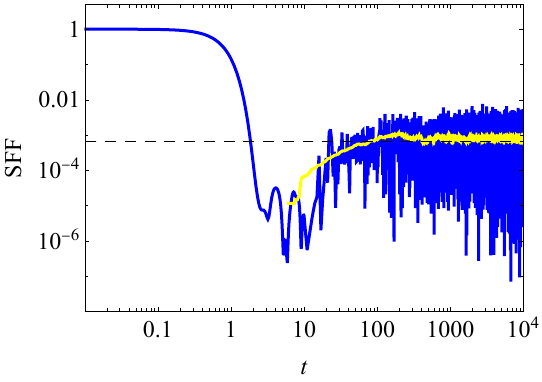} }
     \caption{
     Spectral Form Factor in the NNN deformation of the  XXZ spin chain with $\beta = 0$ and $N=15$ for the positive parity sector $H_1$ with total spin in the $z$-direction $M_z=-5/2$. The dashed lines are \eqref{MFI_SFF_late_time}. The yellow line denotes the moving average.}
     \label{XXZ_SFF}
\end{figure}

In addition, we find that the late-time behavior of SFF for the TFD state at infinite temperature $\beta=0$ is related to the saturation value of spread complexity according to the relation
\begin{align}\label{MFI_SFF_late_time}
   {\lim_{T\rightarrow \infty } \frac{1}{T}\int_{0}^{T}\text{SFF}(t) = \frac{1}{d_i} =\frac{1}{1+2C(t=\infty)} \,,}
\end{align}
where the first equality is derived from \cite{Cotler:2016fpe}, while the second is obtained through \eqref{spread_late_time}~\cite{Erdmenger:2023wjg,Rabinovici:2020ryf,Rabinovici:2022beu}.
This is depicted as dashed lines in Fig.~\ref{MFI_SFF}.\footnote{We make the plot of $1/d_i$ since ~\eqref{spread_late_time} does not hold for systems exhibiting degeneracy within the energy spectrum, as discussed in footnote \ref{footnote7ref}.}

\paragraph{NNN Deformation of the Heisenberg Chain.}
Figs.~\ref{XXZ_Log_spread} and~\ref{XXZ_SFF} display the spread complexity and SFF for the Heisenberg model. 
We find similar features to those in the mixed-field Ising model. The transition time scales of spread complexity closely align with those of the SFF. The presence of a clear ramp in the SFF is related with the occurrence of a peak in spread complexity. Also, the late-time saturation of the SFF adheres to~\eqref{MFI_SFF_late_time}.

\subsection{Spectral Complexity}\label{subsec:ResultsSpectral}

\subsubsection{Early-Time Behavior and Characteristic Timescales}\label{subsubsec:ResultsSpectralTimesc}

In this section we show the results for spectral complexity in spin chain models. We focus on two main aspects: the time-scale at which the saturation occurs and the early-time behavior. 

Regarding saturation, it is worth noting that for the chaotic case, the saturation time-scale of \textit{spread complexity} is of a comparable order to that of their integrable counterparts (refer to Sec.~\ref{subsec:SpreadC}).
However, there are examples in simple quantum systems, such as quantum billiards (\cite{Camargo:2023eev}), where the saturation time-scale of \textit{spectral complexity} for chaotic systems is several orders of magnitude smaller than in their integrable counterparts.\footnote{As advertised in the introduction, in Sec.~\ref{subsubsec:ResultsSpectralRMTTime} we provide an explanation for this disparity from the perspective of RMT at finite size $N$.}

On the contrary, unlike the late-time saturated behaviors of both complexities, it has been suggested~\cite{Erdmenger:2023wjg} that the early-time behaviour of spectral and spread complexity {in the RMT studies} can be associated each other through an Ehrenfest theorem in Krylov space.
We study all these aspects above within the context of our spin chain models.

\paragraph{Mixed-field Ising model.}

\begin{figure}[]
 \centering
     \subfigure[Spectral complexity]
     {\includegraphics[width=6.6cm]{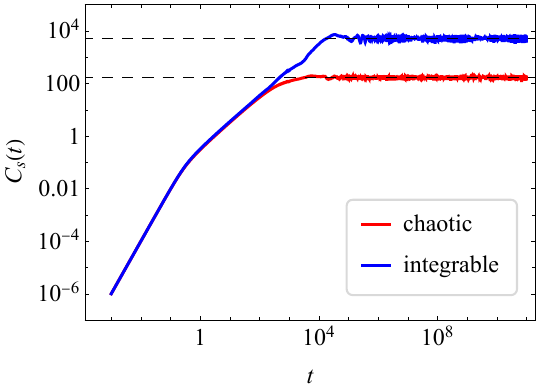} }
     \subfigure[Early time behavior]
     {\includegraphics[width=6.25cm]{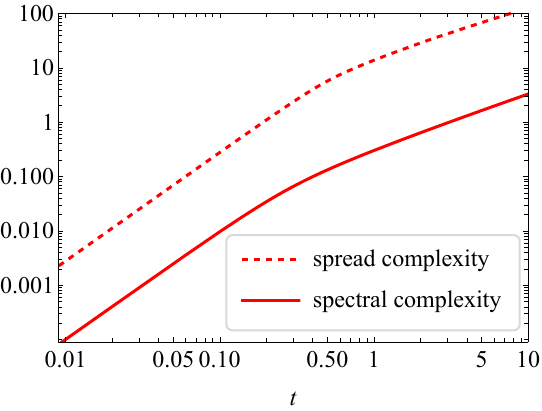} }
     \caption{Spectral complexity in mixed-field Ising model with $H_1$, $\beta = 0$, $N=12$. \textbf{Left:} Spectral complexity with chaotic case (red line) and integrable case (blue line). The black dashed line is saturation value of \eqref{spectral_late_time}. \textbf{Right:} Early time behavior. The red solid line is spectral complexity and red dashed line is spread complexity.}
     \label{MFI_spectral}
\end{figure}

In the left panel of Fig.~\ref{MFI_spectral}, we show the time evolution of spectral complexity. A typical feature is that the spectral complexity saturates at a later time scale for the integrable case (blue line) compared with the chaotic case (red line). This observation supports the observation that spectral complexity is sensitive to signatures of quantum chaos such as energy-level repulsion. In the Figure, the saturation value~\eqref{spectral_late_time} is represented by a black dashed line.

Nevertheless, in the right panel of Fig.~\ref{MFI_spectral}, we find a distinction between spectral complexity (solid line) and spread complexity(dashed line) during early times within the mixed-field Ising model, unlike the findings in the RMT study in \cite{Erdmenger:2023wjg}. We will provide an explanation for this discrepancy below.

\paragraph{NNN Deformation of the Heisenberg Model.}

\begin{figure}[]
 \centering
     \subfigure[Spectral complexity]
     {\includegraphics[width=6.6cm]{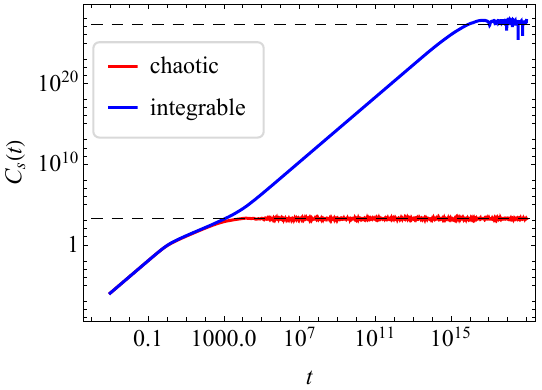} }
     \subfigure[Early time behavior]
     {\includegraphics[width=6.25cm]{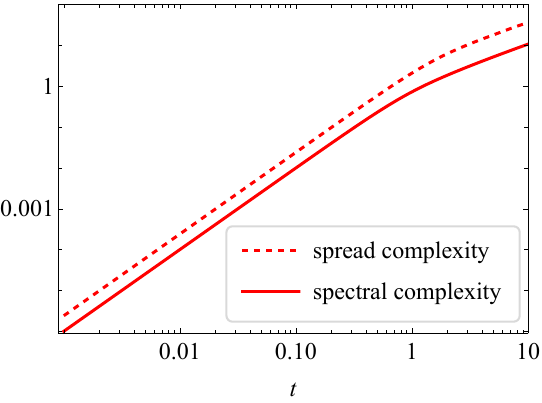} }
     \caption{Spectral complexity in the NNN deformation of the XXZ spin chain with $\beta = 0$ and $N=15$ and total spin in the $z$-direction $M_z=-5/2$ for the positive parity sector $H_1$. \textbf{Left:} Spectral complexity for the chaotic (red line) and integrable (blue line) cases. The black dashed line is the saturation value, according to~\eqref{spectral_late_time}. \textbf{Right:} Early-time behavior comparison between spectral (solid line) and spread (dashed line) complexity.}
     \label{XXZ_spectral}
\end{figure}

Likewise, we display the spectral complexity of the next-nearest-neighbor (NNN) deformation of the Heisenberg model in Fig.~\ref{XXZ_spectral} and observe similar outcomes. Notably, the saturation time-scale for the integrable case significantly lags behind that of the chaotic case. Additionally, we observe a discrepancy between spectral and spread complexity in the early-time regime.

Now, we discuss the discrepancy between spectral and spread complexity in the early-time regime within our spin chain models when $\beta=0$. To explore this, it is beneficial to examine the early-time behavior of both \eqref{spread complexity} and \eqref{eq:SpectralComp}, yielding:
\begin{align}\label{EarlyBehavior}
\begin{split}
C(t) \approx b_{1}^2 \, t^2 \,, \qquad
C_S(t) \approx \left( 1-\frac{1}{d}\right) t^2 \,,
\end{split}
\end{align}
 where the former has been derived in \cite{Huh:2023jxt} for generic quantum systems by analytically solving the equation \eqref{Lanczos sch} perturbatively in the early time. Then, it becomes evident that, in general, the two complexities behave differently unless $b_1=1$ and $d\gg1$. We find within our spin chain models that $b_1\neq1$:
this is the primary reason for the disparities in complexities during the early time regime. It is noteworthy that in the RMT studies in \cite{Erdmenger:2023wjg} when the dimension of Krylov space is large enough, $b_1=1$.

The Ehrenfest theorem on the relation between $C(t)$ and $C_S(t)$ at early times was formulated for RMTs in the $\tilde{\beta}-$ensemble. The discrepancy in our case could be attributed to the fact that the spectral density of our spin models is not exactly the semi-circle law. Thus, the Ehrenfest theorem for RMTs does not directly apply to our case. For the SYK model, a similar discrepancy was discussed in \cite{Erdmenger:2023wjg}.

\subsubsection{Lessons from Random Matrices: Minimum Energy Difference}\label{subsubsec:ResultsSpectralRMTTime}

In this section, we calculate the minimum energy difference in RMTs. The rationale for examining it is as follows. To derive the long-time average (\ref{spectral_late_time}), we used
\begin{align}
\lim_{T\to\infty}\frac{1}{T}\int_0^T \textrm{d}t  \sin^2 \left(\frac{t\Delta E}{2}\right) \quad\longrightarrow\quad \frac{1}{2} \,,
\end{align}
where $\Delta E=E_m-E_n$ is the energy difference. 
One can also find the time-scale for the saturation as
\begin{align}
\left.\left[\frac{1}{T}\int_0^T \textrm{d}t  \sin^2 \left(\frac{t\Delta E}{2}\right)\right]\right|_{T=\frac{\pi}{\Delta E}} \quad=\quad \frac{1}{2} \,.
\end{align}
Since the definition of spectral complexity (\ref{eq:SpectralComp}) contains $\frac{1}{(E_{m}-E_{n})^2}$, the term with the smallest $|E_m-E_n|$ would contribute dominantly to the saturation value of $C_S(t)$.
For the long-time average, we should take a longer time than $\frac{\pi}{\Delta E_{\text{min}}}$, where $\Delta E_{\text{min}}:=\text{min}\{|E_m-E_n|\}$ is the minimum energy difference in quantum systems. Therefore, $\Delta E_{\text{min}}$ is an important inverse time scale for the saturation of spectral complexity. We study how $\Delta E_{\text{min}}$ depends on size $d$ of Hamiltonian matrix $H$.

We consider three famous random matrix ensembles of $d\times d$ matrices: Gaussian Orthogonal Ensemble (GOE), Gaussian Unitary Ensemble (GUE), and Gaussian Symplectic Ensemble (GSE). Our convention of Gaussian measures is
\begin{align}\label{GaussianMeasures}
\frac{1}{Z_{\text{GOE}}}e^{-\frac{d}{4}\Tr H^2}, \;\;\;\;\;\; \frac{1}{Z_{\text{GUE}}}e^{-\frac{d}{2}\Tr H^2}, \;\;\;\;\;\;\frac{1}{Z_{\text{GSE}}}e^{-d \,\Tr H^2},
\end{align}
respectively, where $H$ are $d\times d$ matrices with symmetry of each ensemble, and $Z_{\cdots}$ are normalization constants. With this convention, the spectral density of RMTs at large $d$ is given by the Wigner semicircle distribution with radius $R=2$, which is $d$-independent.

Since energy eigenvalues of RMTs are random variables, one needs to take some kind of average. We take an average of the minimum energy difference by iteratively generating $d$ energy eigenvalues from the Gaussian measures (\ref{GaussianMeasures}) \cite{Caputa:2024vrn}. Specifically, we generate $\mathcal{N}_{\text{it}}$ sets of $d$ energy eigenvalues and define the average of minimum energy difference $\overline{\Delta E_{\text{min}}}$ by
\begin{align}
\overline{\Delta E_{\text{min}}}:=\frac{1}{\mathcal{N}_{\text{it}}}\sum_{i=1}^{\mathcal{N}_{\text{it}}}\Delta E_{\text{min}}^{(i)},
\end{align}
where $\Delta E_{\text{min}}^{(i)}$ is the minimum energy difference in the $i$-th set of $d$ energy eigenvalues. Note that, the energy spectrum in GSE is doubly degenerate, and thus we take $\Delta E_{\text{min}}^{(i)}$ in GSE to be the minimum energy difference between non-degenerate energies.

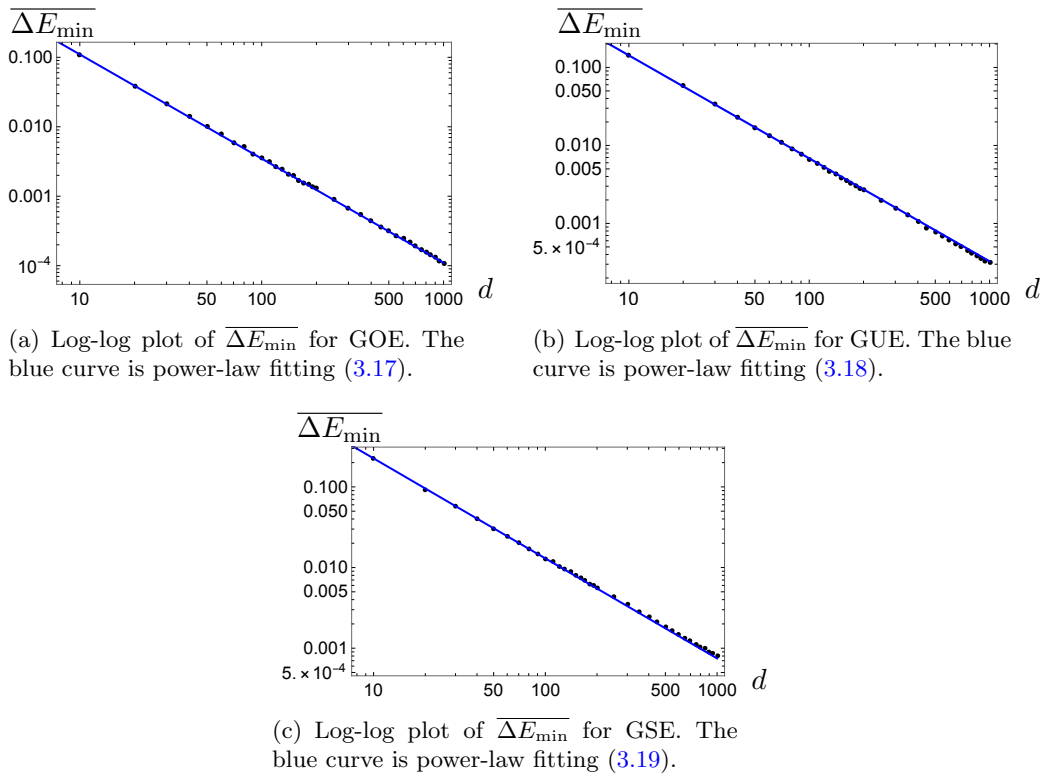
\begin{figure}[]
 \centering
     \subfigure[Log-log plot  of $\overline{\Delta E_{\text{min}}}$ for GOE. The blue curve is power-law fitting (\ref{GOEfit}).]
     {\includegraphics[width=6 cm]{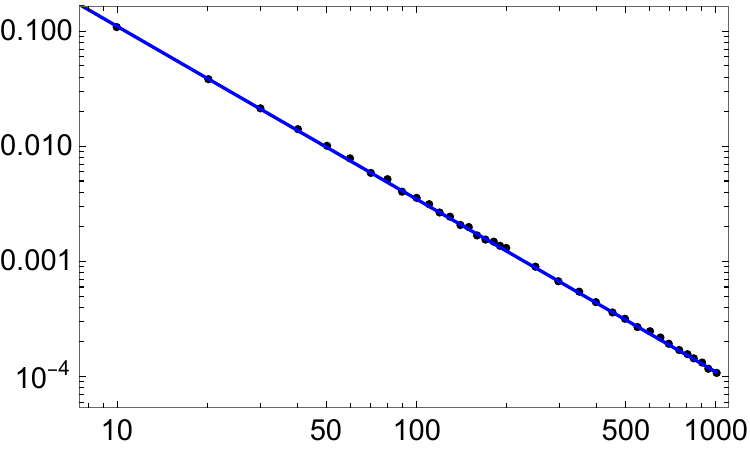}
          \put(-170,105){$\overline{\Delta E_{\text{min}}}$}
     \put(5,5){$d$}
     }
     \qquad
 \centering
     \subfigure[Log-log plot of $\overline{\Delta E_{\text{min}}}$ for GUE. The blue curve is power-law fitting (\ref{GUEfit}).]
     {\includegraphics[width=6.3 cm]{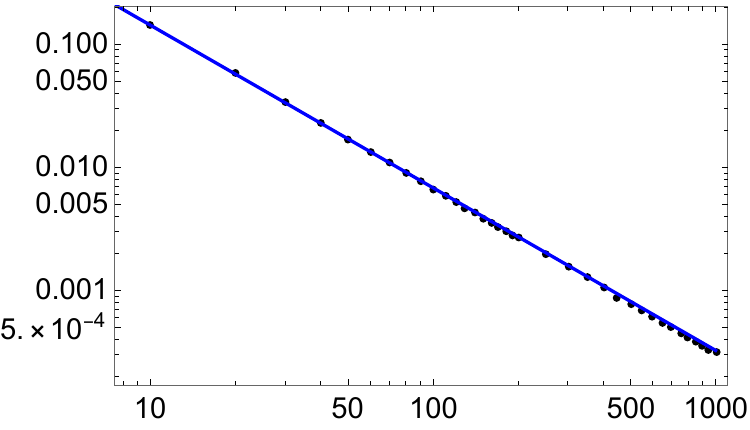}
            \put(-170,105){$\overline{\Delta E_{\text{min}}}$}
        \put(5,5){$d$}}
 \centering
         \subfigure[Log-log plot of $\overline{\Delta E_{\text{min}}}$ for GSE. The blue curve is power-law fitting (\ref{GSEfit}).]
        { \includegraphics[width=6 cm]{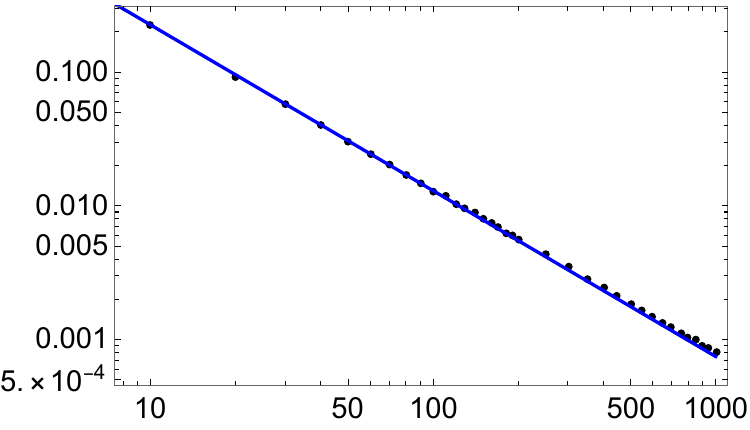}
            \put(-165,100){$\overline{\Delta E_{\text{min}}}$}
        \put(5,5){$d$}}
      \caption{$d$-dependence of $\overline{\Delta E_{\text{min}}}$ in RMTs.}
     \label{fig:EminRMT}
\end{figure}

Fig. \ref{fig:EminRMT} shows log-log plots of $\overline{\Delta E_{\text{min}}}$, where we choose $\mathcal{N}_{\text{it}}=1000$ for the average. The black dots are the averaged values of our numerical data generated by the Gaussian measures (\ref{GaussianMeasures}), and the blue curves are power-law fitting of the $d$-dependence given by
\begin{align}
\overline{\Delta E_{\text{min}}} \,\approx&\,\,\frac{3.48}{d^{1.50}} \;\;\; (\text{GOE}),\label{GOEfit}\\
\overline{\Delta E_{\text{min}}} \,\approx&\,\,\frac{2.98}{d^{1.32}} \;\;\; (\text{GUE}),\label{GUEfit}\\
\overline{\Delta E_{\text{min}}} \,\approx&\,\,\frac{3.92}{d^{1.24}} \;\;\; (\text{GSE}).\label{GSEfit}
\end{align}
Although there will be numerical errors due to random variables, the $d$-dependence of $\overline{\Delta E_{\text{min}}}$ can be well fitted by power-law behaviors as seen in linear behaviors in the log-log plots.

\begin{figure}[]
 \centering
     \subfigure[Positive parity sector]
     {\includegraphics[width=6 cm]{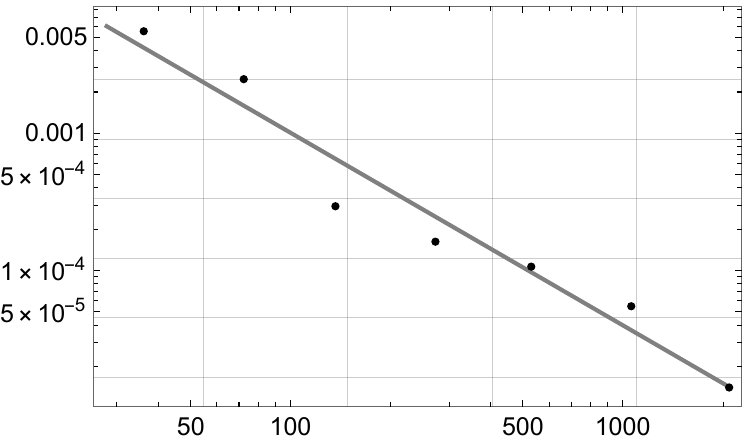}
     \put(-10,0){\small $d_1$}
     \put(165,0){\small $d_2$}
     \put(-165,105){\small $\Delta E_\text{min}$}
     \put(10,105){\small $\Delta E_\text{min}$}
     }
     \subfigure[Negative parity sector]
     {\includegraphics[width=6 cm]{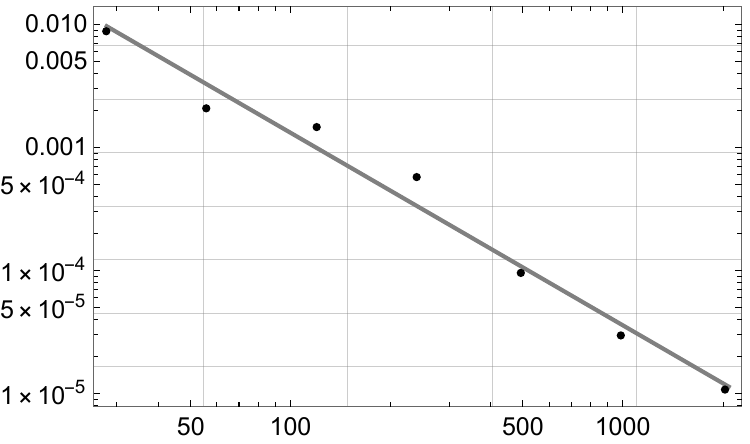}}
      \caption{Minimum energy difference versus $d_1$ or $d_2$ (see Fig.~\ref{symmetry}) for the mixed-field Ising model in the chaotic case ($J=1$, $h_x=-1.05$, and $h_z=0.5$). Here, to better compare with the results for random matrices, we re-scaled the spectrum of the system to make it fit between 0 and 2, i.e., $[E_\text{min},E_\text{max}]=[0,2]$. The fitted lines are $\Delta E_\text{min}=0.65/(d_1)^{1.40}$ for the even parity sector and $\Delta E_\text{min}=1.78/(d_2)^{1.57}$ for the odd parity sector. This power-law behavior does not happen in the integrable cases. }
     \label{fig:MinEnergyIsingChaotic}
\end{figure}

\paragraph{Results for the mixed-field Ising Model.}
The analysis for random matrix ensembles reveals that the minimum energy difference displays a power-law behavior as a function of the size $d$ of the matrix ($\Delta E_\text{min} = \frac{a}{d^b}$, for some constants $a$ and $b$). Since strongly chaotic systems are expected to display random matrix behavior, we expect that the minimum energy difference also display such power-law behavior in these cases. To further investigate this, we study the minimum energy difference versus the size of the matrices in each symmetry sector of the mixed-field Ising model. In the chaotic case, we find the expected power-law behavior: 
\begin{equation} \label{sc123}
   \Delta E_\text{min}=0.65/(d_1)^{1.40} \,, \qquad \Delta E_\text{min}=1.78/(d_2)^{1.57}
\end{equation}
for the even parity sector and the odd parity sector respectively. See Fig.~\ref{fig:MinEnergyIsingChaotic}. The results seem closer to the behavior observed for GOE, which describes systems with time-reversal symmetry as is the case for the mixed-field Ising model. By contrast, in the integrable case, we did not observe this power-law behavior. {The level spacing distribution in the integrable case is the Poisson distribution as seen in Fig.~\ref{NNSD}. Due to nonzero distribution at $s=0$, the minimum energy difference in the integrable case can be erratic compared to the power-law behavior.}

Since the saturation time of spectral complexity is given roughly by $t_\text{saturation} \sim 1/\Delta E_{\text{min}}$, the above-mentioned power-law behavior implies the following scaling behavior of the saturation time in the chaotic case
\begin{equation}
    t_\text{saturation} \sim d^b = e^{b \log d}\,,
\end{equation}
which (roughly) agrees with the expected behavior for chaotic systems proposed in \cite{Iliesiu:2021ari}, namely, $t_\text{saturation} \sim e^S$,  if we identify $\log d$ as the entropy $S$ of the system. 

\section{Discussion and Conclusions}\label{sec:Discussion}

In this work, we studied two different measures of quantum complexity in the context of spin systems which interpolate between integrable and chaotic, according to their spectral statistics. First, we showed that whenever the systems are in their chaotic point, the spread complexity of the Thermofield Double (TFD) state presents a peak at a timescale which is comparable to the one in which the spectral form factor (SFF) transitions from the ramp to a plateau, strengthening the observations made in~\cite{Balasubramanian:2022tpr,Erdmenger:2023wjg} that the presence of such a feature could be used to signal chaotic quantum dynamics.

It is worth noting that the authors of~\cite{Gautam:2023bcm} also examined spread complexity in spin systems capable of exhibiting chaotic behavior. Nevertheless, no distinct peak was observed in their findings (though an analysis of the Lanczos coefficients pointed towards chaotic behaviour). One plausible explanation for this discrepancy can be traced back to the original work on spread complexity~\cite{Balasubramanian:2022tpr}. In that study, the spread complexity conjecture for quantum chaos was formulated based for the TFD state, i.e., an entangled state. Therefore, the lack of a distinct peak in~\cite{Gautam:2023bcm} might be attributed to a different choice of initial state. To be precise, one important feature that distinguishes the TFD state from the domain wall state $\vert \!\uparrow \cdots \uparrow \downarrow \cdots \downarrow \rangle$ considered in that work is the following: while the former has a non-zero overlap with energy eigenstates with arbitrary energies, it is more ``sensitive'' to the complete energy spectrum, the latter is a very specific state with a more restricted overlap with energy eigenstates. Since the presence of a peak can be attributed to spectral rigidity, requiring access to arbitrary states with arbitrary energy, the TFD state is more suitable for studying such effects.\footnote{We thank Kuntal Pal for pointing out this possibility.} Moreover, the authors of~\cite{Scialchi:2023bmw} studied the spread complexity and the Lanczos coefficients for the mixed-field Ising model for different choices of states. They show that the saturation value of spread complexity as well as the variance of the Lanczos coefficients can probe chaotic behavior for certain choices of states. However they do not discuss the choice of the TFD state as the initial state or the peak in spread complexity.

Similarly, we find that the saturation value of spread complexity after the peak depends on details of the dynamics that may not be attributed solely to the spectral statistics. While we do not have a complete understanding of this feature, this difference can potentially be attributed to the choice of initial state and its spreading over the Krylov basis. To be precise, since spread complexity measures how much the initial states spread through the Krylov basis, the fraction of the Krylov basis that is occupied in the chaotic case at late times is generically larger than the fraction of the Krylov basis that is occupied integrable case. However, the precise amount of spreading that the initial state has with the Krylov basis depends on the properties of the system beyond its spectral properties in a non-trivial way.

For the TFD state at finite temperature $1/\beta$, we found that the feature of the peak becomes more visible as the temperature increases. This is consistent with the expectation that increasing the temperature enables us to access states with higher energy, which is necessary to accurately capture the spectral statistics of the theory.

It is worth pointing out that the time-dependent properties of other quantum complexity measures, such as time-evolution of Nielsen complexity, have also been studied in the context of integrable and chaotic systems. In particular, a bound on Nielsen complexity has been efficiently estimated for bi-invariant metrics using lattice cryptography techniques~\cite{Craps:2022ese,Craps:2023rur}. Building on these results, authors in~\cite{Craps:2023ivc} furthermore uncovered a relation between the time average of spread complexity and a bound on Nielsen complexity. It would be interesting to continue to explore the relation between these two complexity measures and understand whether there exist bounds to the growth rate of spread complexity akin to~\cite{Hornedal:2022pkc}, that could help understand the physics behind its saturation value, such as its state dependence.

We also studied spectral complexity and provided further evidence to the observation made in~\cite{Camargo:2023eev} that its saturation value and timescale for the chaotic point of a given Hamiltonian that interpolates between integrable and chaotic, is orders of magnitude smaller than for the integrable point. By analyzing the properties of spectral complexity for random matrix theories (RMTs) of finite size $d=2^N$, we determined that the property of the system determining these features is the minimum of the energy difference in the spectrum. To the best of our knowledge, this is the first time that such a feature of the spectrum appeared in the context of quantum chaos.

\paragraph{Timescales of dynamical probes of quantum chaos.}
Perhaps the most accepted definition of quantum chaos is based on the idea that Hamiltonians of chaotic systems behave like random matrices. The basic idea is that the energy levels of the system in each symmetry sector are strongly correlated, and their nearest-neighbor level spacing statistics follows a Wigner--Dyson distribution. By contrast, the energy levels of integrable systems are uncorrelated and their nearest-neighbor level spacing statistics typically follows a Poisson distribution. This definition is based on the Bohigas--Giannoni--Schmit (BGS) conjecture \cite{Bohigas:1983er}, which states that certain fluctuation properties of the spectrum of time-reversal classically chaotic systems (K-systems) are well described by Gaussian orthogonal ensembles (GOE) of random matrices.  This conjecture is still not proven but is supported by a vast number of numerical studies, with well-understood limitations in its validity. In the case of systems that do not have a classical limit, it is natural to use the random matrix behavior as a working definition of quantum chaos. This definition is very economical since it only depends on the Hamiltonian of the system. Nevertheless, in addition to studying the spectrum of the systems, which are {\it static}, exploring {\it time-dependent} quantities can also offer additional valuable insights for understanding manifestations of chaos at the quantum level. In particular, it is interesting to investigate how the chaotic properties of the Hamiltonian influence the dynamics of states and operators and what are the \emph{signatures} of chaos for them. In Table. \ref{tab2}, including our findings on spread and spectral complexities, we summarize some well-studied probes of quantum chaos and their characteristic timescales.


\begin{table}[]
\centering
\begin{tabular}{ |c|c| }
\hline
 \textbf{Probes} & \textbf{Characteristic Timescales}   \\
\hline
\hline
Out-of-time-order correlator &  $t_\text{scrambling} \sim \log d$ ~\cite{Sekino:2008he, Gharibyan:2018jrp} \\
\hline
Krylov operator complexity &  $t_\text{exponential-growth} \,\sim\, \log d $\,\,, $t_\text{saturation} \sim e^{2d}$ \cite{Rabinovici:2022beu} \\
\hline
Spectral Form Factor & $t_\text{plateau} \,\sim\, t_H = d$ ~\cite{Cotler:2016fpe}   \\
\hline
Spread complexity & $t_\text{peak} \,\lesssim\, t_{H} = d$   \\
\hline
Spectral complexity & $t_\text{saturation} \,\sim\, d^{\,b}$, with $b$ given by Eqs. (\ref{GOEfit}-\ref{GSEfit}, \ref{sc123})  \\
\hline
\end{tabular}
\caption{Typical timescales characterizing different time-dependent probes of quantum chaos. Here, $d$ is the dimension of the Hilbert space, $N$ is the number of lattice sites, and $t_H$ is the Heisenberg time.}\label{tab2}
\end{table}

The standard time-dependent quantity in studies of quantum chaos is the Spectral Form Factor (SFF). In RMTs, SFF as a function of time displays a typical slope-dip-ramp-plateau behavior, characterized by certain timescales. The same behavior is expected to occur in chaotic systems, while in integrable systems one expects a different behavior. Spread and spectral complexities provide a new window into this time dynamics.
\begin{figure}[]
\centering
\subfigure[Spectral form factor]
     {\includegraphics[width=7cm]{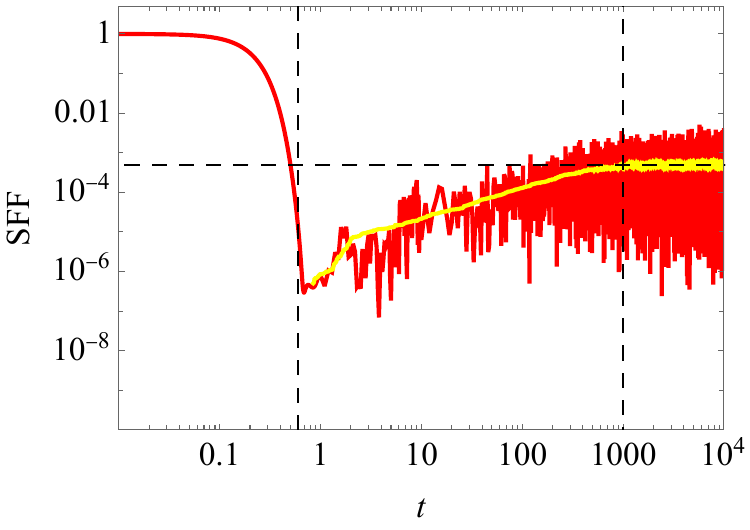}
     \includegraphics[width=7cm]{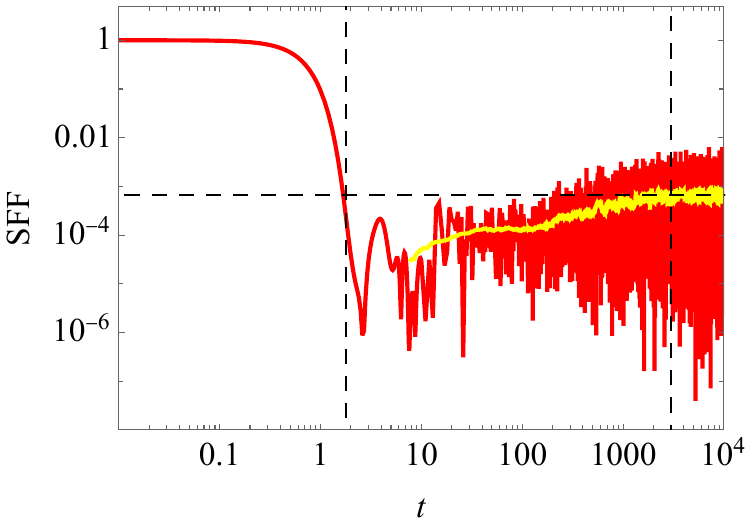}}
\subfigure[Spread complexity]
     {\includegraphics[width=7cm]{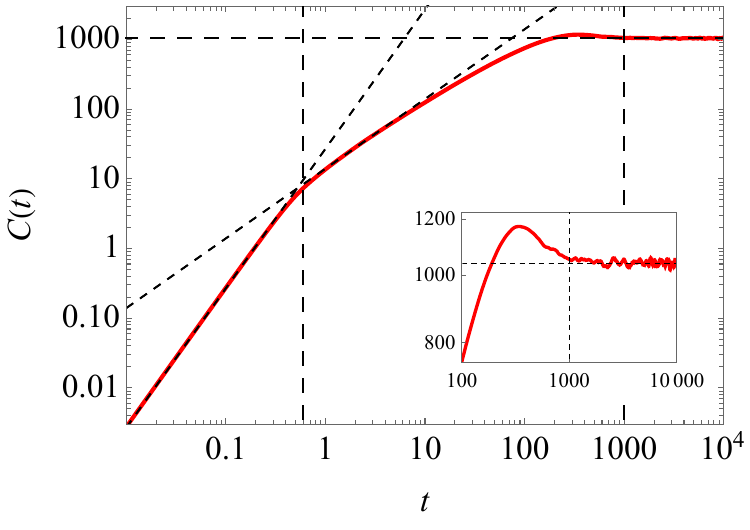}
    \includegraphics[width=7cm]{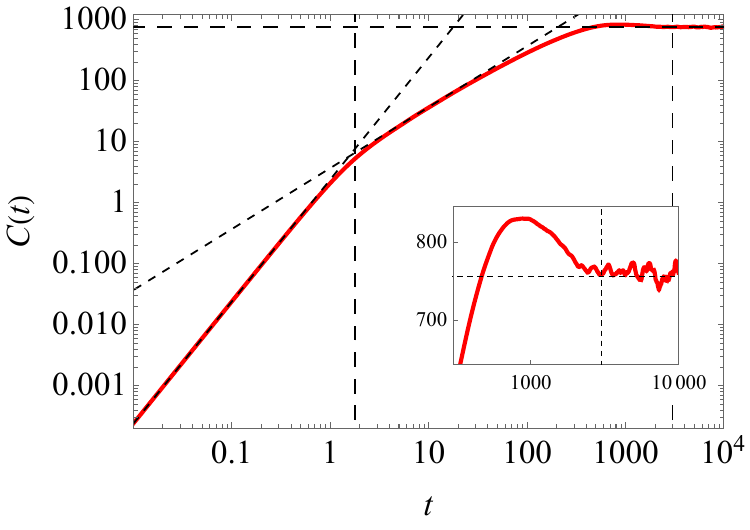}}
\subfigure[Spectral complexity]
     {\includegraphics[width=7cm]{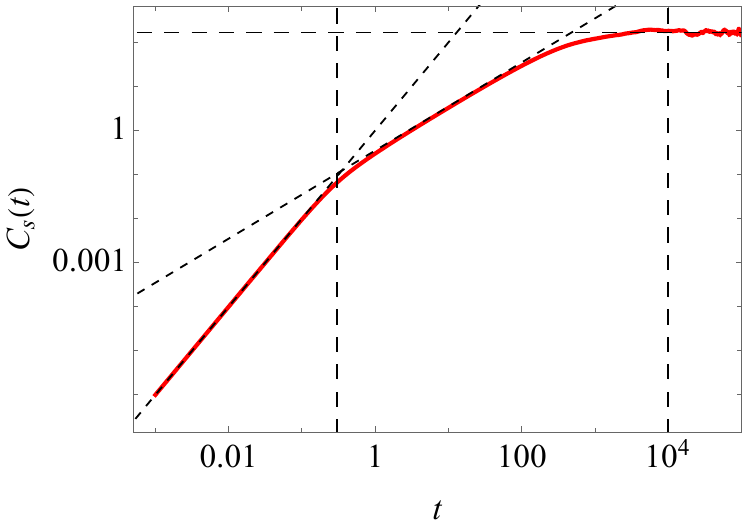}
     \includegraphics[width=7cm]{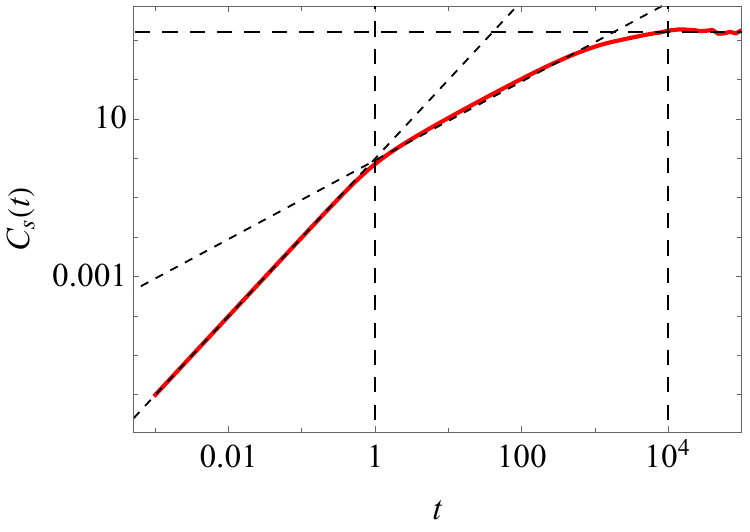}}
\caption{ Comparison of the characteristic timescales for the mixed-field Ising model with $d_1 = 2080$ (left) and the deformed Heisenberg XXZ model with $d_1 = 1512$ (right) for the chaotic cases. In both cases we consider the positive parity sector. Additionally, in the deformed Heisenberg model we consider the sector with $M_z=-5/2$. \label{fig:timescale-comparison}}
\end{figure}
In Fig. \ref{fig:timescale-comparison}, we present a summary of our findings regarding the timescales on SFF, spread complexity, and spectral complexity, for the chaotic cases and positive parity sectors of the mixed-field Ising model and the deformed XXZ model. 

The behavior of SFF is characterized by two distinct timescales, $t_\text{dip}$ and $t_\text{plateau}$, denoted as the vertical dashed lines in the top panel of Fig. \ref{fig:timescale-comparison}. These timescales can be associated with the timescales of spread and spectral complexities as follows.

Initially ($t \lesssim t_\text{dip}$), both spread and spectral complexities exhibit quadratic growth over time, while the SFF displays a discernible slope. As time progresses ($t \gtrsim t_\text{dip}$), spread complexity and spectral complexity transition to an apparent\footnote{More precisely, the spectral complexity appears to display a linear growth in our log-log plots, but its behavior is more complicated. Its second derivative is very small but nonzero, oscillating considerably in this time range. This should be the case since the second derivative of the spectral complexity gives the SFF.} linear growth, and the SFF begins oscillating significantly, although on average it displays a linear ramp-like trend.
Then, around $t \sim t_\text{plateau}$, both spread complexity and the SFF reach saturation, with spread complexity peaking shortly before saturation. 
Spectral complexity, on the other hand, saturates at a slightly later timescale, denoted as $t_\text{saturation}$: we observe that $t_\text{dip} \lesssim 1$, while $t_\text{plateau} \sim d_i$ and $t_\text{saturation} \sim d_i^{b}$, with $b$ given by given by Eqs. (\ref{GOEfit}-\ref{GSEfit}, \ref{sc123}).

It is notable that we also checked that two timescales ($t_\text{dip}$ and $t_\text{plateau}$) are comparable between SFF and spread complexity even for the integrable cases. This indicates that such timescales are characteristics of the system's symmetry and dimension rather than chaos. Instead of the timescale, the presence of a peak in spread complexity and a linear ramp in the SFF serve as signatures of chaos. The saturation values of spread complexity and the SFF vary depending on the level of chaos and the system's state. However, for the TFD state at infinite temperature, these saturation values are maximal and determined by the system's symmetry and dimension, as demonstrated in equation \eqref{MFI_SFF_late_time}. As for spectral complexity, both the saturation timescale $t_\text{saturation}$ and late-time values serve as signatures of chaos due to differences in the level spacing distribution.

Last but not least, regarding the SFF, it is also notable that our observation of $t_\text{dip} \sim \mathcal{O}(1)$, regardless of the system size (when $6\leq N \leq12$), stands in contrast to the behavior identified in RMTs, where $t_\text{dip} \sim d^{1/2}$ \cite{Cotler:2016fpe}. In RMT, this scaling of the dip time is influenced by the disconnected part of the SFF, primarily affected by the region near the spectrum's edge, characterized by a semicircular density of states. However, in the case of spin chains, we have identified a Gaussian-shaped density of states. The deviation in scaling behavior from RMT can thus be attributed to this disparity in the density of states exhibited by the spin chain under examination. However, it remains unclear to us why the Gaussian density of states shape results in a dip time seemingly independent of the system size. We are currently investigating this behavior.

It is worth recalling that these state-dependent (spread and spectral) complexities can probe random matrix behavior well when the complexities are associated with the TFD state. This happens because random matrix behavior is a statement about the full\footnote{One should note, however, that random matrix behavior is normally found excluding the edges of the energy spectrum of the system, where the behavior is typically non-universal. Certain systems, however, display chaotic fluctuations through the entire spectrum~\cite{Altland:2024ubs}.} (level-spacing) spectrum of the system. As such, in this regard, any quantity that aims at probing random matrix behavior would benefit from having information about the full spectrum of the system, and the TFD state accomplishes this. We believe that other states sufficiently similar to the TFD state are also capable of effectively probing random matrix behavior.

Other notions of quantum chaos, such as out-of-time-order correlators (OTOCs) and Krylov operator complexity, measure the scrambling of quantum information, and are inherently associated with states. Typically, they are computed from correlation functions of operators in a thermal state. This parallels classical ergodic theory, where chaos is often characterized by vanishing of correlation functions (mixing), which are defined with reference to some measure. The measure is commonly interpreted as a state.

\paragraph{Why is the TFD state special for state-dependent probes of quantum chaos?}
In the following, we further discuss why the TFD state seems to be well-suited for detecting random matrix behavior for state-dependent probes of chaos, in particular, for spread complexity. The starting point for computing spread complexity in an arbitrary state $| \Psi \rangle$ is the autocorrelation function
\begin{equation} \label{eqqq1}
    \psi_0(t)=\langle \Psi(0)| \Psi(t) \rangle.
\end{equation}
If $| \Psi \rangle =| \text{TFD} \rangle$, then the autocorrelation function can be written \cite{Balasubramanian:2022tpr,Caputa:2024vrn} as
\begin{equation} \label{eqqq2}
    \psi_0(t)= \frac{Z_{\beta-i t}}{Z_{\beta}} = \sqrt{\textrm{SFF}(t)}\,,
\end{equation}
which depends on the full spectrum of the system due to the following expression of SFF
\begin{equation}
    \textrm{SFF}(t):=\frac{\vert Z(\beta+it)\vert^2}{\vert Z(\beta)\vert^2}=\frac{1}{Z(\beta)^2}\sum_{m,n}e^{-\beta(E_{m}+E_{n})}e^{i(E_{m}-E_{n})t}~.
\end{equation}
Since the SFF depends on all energy differences $E_n-E_m$, it is intuitive to consider that this quantity is sensitive to the level spacing statistics of the system. Consequently, the autocorrelation function is also sensitive to the level spacing statistics and this dependence passes to the other wave functions $\psi_n(t)$ with $n>0$, and to the spread complexity $C(t)=\sum_n n \vert\psi_n(t)\vert^2$.

Note that for an arbitrary state $| \Psi \rangle$ in a single copy of the system, the autocorrelation function can be written as
\begin{equation}
    \psi_0(t)=\sum_n e^{i E_n t} \langle \Psi(0)|n \rangle \langle n| \Psi(0) \rangle\,,
\end{equation}
which in general depends on the overlaps between the evolving states and the eigenstates and may have a much simpler dependence on the spectrum $\{ E_n\}$ than the TFD state, being much less sensitive to the level spacing statistics of the system. In particular, if the initial state is an eigenstate of the Hamiltonian, $| \Psi \rangle = |n'\rangle$, its time evolution is simply given by $| \Psi(t) \rangle = e^{i H t}|n'\rangle = e^{i E_{n'} t}|n'\rangle$. The autocorrelation function in this case is determined by a single energy level, not having information about the rest of the spectrum. In this case, the autocorrelation function and spread complexity are insensitive to the full level-spacing statistics of the system. 
This situation is similar to a common understanding that we should not use operators having a zero commutator with the Hamiltonian for OTOCs and Krylov operator complexity as probes of quantum chaos.

The above considerations allows us to conclude that states which are similar to the TFD state will probably be sensitive to the random matrix behavior of the system, while linear combinations of a few energy eigenstates will probably be insensitive to it.
Moreover, when opting for the eigenstate as the initial state:
\begin{align}
|K_0\rangle = |\psi(0) \rangle = | n' \rangle\,, 
\end{align}
it becomes apparent that the Lanczos algorithm \eqref{Lanczos algorithm} fails to yield a non-trivial Krylov basis. This is because eigenvectors of the Hamiltonian have a trivial time evolution, controlled by a simple phase -- the initial state, which is given by $|K_0\rangle$, does not spread through the rest of the Krylov basis $|K_n\rangle$ for $n>0$. The first vector of the Krylov basis, $| K_1 \rangle$, vanishes due to
\begin{align}\label{}
\begin{split}
| A_{1} \rangle 
&= (H - a_0) | K_0 \rangle = (H - n') | n' \rangle = 0 \,,
\end{split}
\end{align}
where we used \eqref{forconclu} in the second equality.
Therefore, employing the Lanczos algorithm reveals the vanishing of all Krylov bases $|K_n\rangle$  for $n>0$, and the corresponding spread complexity vanishes identically, failing to probe the statistical properties of the spectrum. 

\paragraph{Concluding remarks.}
Excluding the cases of saddle-dominating scrambling, the relationship between random matrix behavior and scrambling behavior appears interconnected yet not identical. It is intriguing to explore whether one of these concepts implies the other, potentially serving as a more fundamental notion. We speculate that a precisely defined mathematical definition for quantum chaos, encompassing a quantum version of the ergodic hierarchies, would probably clarify the state-dependent issue of quantum chaos. To the best of our knowledge, this is not well understood yet and remains an area of ongoing exploration.\footnote{For recent proposals of a quantum version of the classical ergodic hierarchies, we refer to \cite{Gesteau:2023rrx, Ouseph:2023juq}.}

The TFD states are well-studied from the perspective of holography because two-sided black holes can be regarded as the holographic duals of TFD states. The authors of \cite{Rabinovici:2023yex} show that \textit{spread} complexity of chord states in the triple scaling limit of the SYK model, which captures the early time behavior of the double-scaled SYK model, matches the Lorentzian wormhole length in Jackiw-Teitelboim gravity. If this duality remains valid for the double-scaled SYK model at late times, then perhaps one should observe a peak in the plot of the quantum-corrected wormhole length. On the other hand, the authors of \cite{Iliesiu:2021ari} proposed that the quantum-corrected wormhole length should be dual to \textit{spectral} complexity, which in general is different from spread complexity -- for example, spectral complexity does not show a peak in our calculations for spin chains. It would be beneficial to elucidate and clarify this apparent inconsistency in subsequent research endeavors.

\acknowledgments
We would like to thank {Kuntal Pal, Juan F. Pedraza, Le-Chen Qu} for valuable discussions and correspondence.  
This work was supported by the Basic Science Research Program through the National Research Foundation of Korea (NRF) funded by the Ministry of Science, ICT $\&$ Future Planning (NRF-2021R1A2C1006791) and the AI-based GIST Research Scientist Project grant funded by the GIST in 2014.
This work was also supported by Creation of the Quantum Information Science R$\&$D Ecosystem (Grant No. 2022M3H3A106307411) through the National Research Foundation of Korea (NRF) funded by the Korean government (Ministry of Science and ICT).
H.-S Jeong acknowledges the support of the Spanish MINECO ``Centro de Excelencia Severo Ochoa'' Programme under grant SEV-2012-0249. This work is supported through the grants CEX2020-001007-S and PID2021-123017NB-I00, funded by MCIN/AEI/10.13039/501100011033 and by ERDF A way of making Europe.
H.~A. Camargo, V.~Jahnke and M.~Nishida were supported by the Basic Science Research Program through the National Research Foundation of Korea (NRF) funded by the Ministry of Education (NRF-2022R1I1A1A01070589, RS-2023-00248186, RS-2023-00245035).
All authors contributed equally to this paper and should be considered as co-first authors.

\appendix
\section{Mixed-field Ising model without considering symmetry sectors}\label{AppA}

\begin{figure}[]
 \centering
     \subfigure[$(h_x,h_z) = (-1.05,0.5)$]
     {\includegraphics[width=6.6cm]{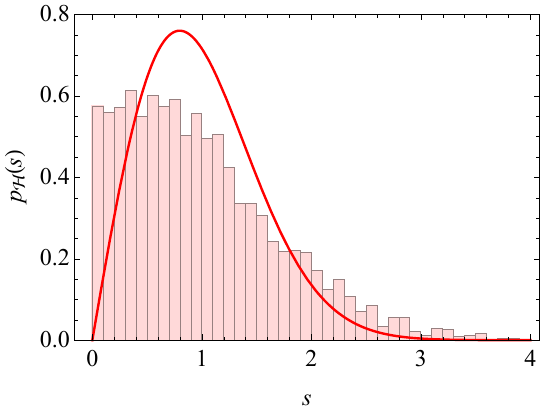} }
     \subfigure[$(h_x,h_z) = (-1,0.001)$]
     {\includegraphics[width=6.6cm]{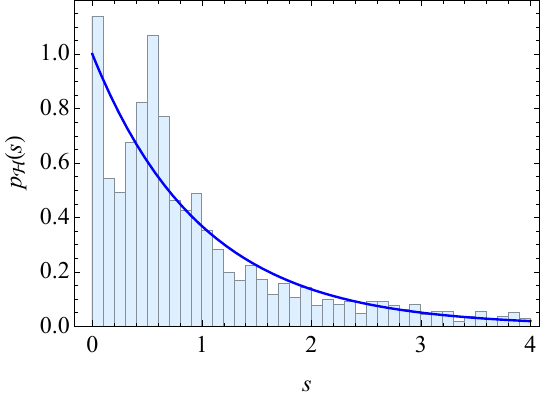} }
     \caption{Level spacing distribution in the mixed-field Ising model without considering the parity sectors (i.e., using the full Hamiltonian $H$) with $N=12$ sites. The red line denotes the Wigner--Dyson distribution (GOE) and the blue line denotes the Poisson distribution.}
     \label{NNSD0}
\end{figure}

\begin{figure}[]
 \centering
     {\includegraphics[width=6.6cm]{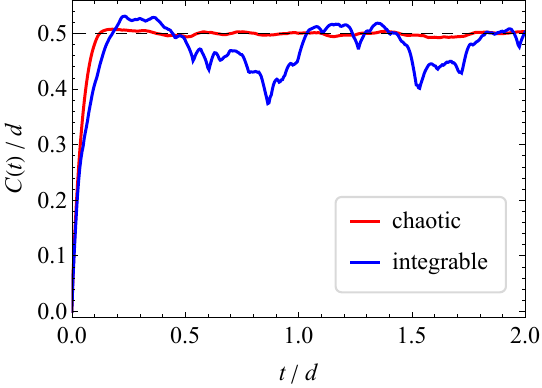} }
     \caption{Spread complexity in the mixed-field Ising model without symmetry block ($H$) for $N=12$, where $d=d_1+d_2$. The red line denotes the chaotic case and the blue line denotes the integrable case. The black dashed line is the asymptotic value \eqref{spread_late_time} of spread complexity at late times.}
     \label{MFI_spread_N12_0}
\end{figure}

In this section, for pedagogical purposes, we present the results of the mixed-field Ising model devoid of symmetry sector considerations. Our analysis explicitly reveals that not considering symmetry sectors diminishes the prominence of the features characterizing chaotic behavior. Specifically, a comparison between Fig. \ref{NNSD} and Fig. \ref{NNSD0} regarding the level spacing distribution, as well as Fig. \ref{MFI_spread_N12} and Fig. \ref{MFI_spread_N12_0} concerning spread complexity, elucidates this observation.

The reason behind this can be understood as follows: In chaotic systems, energy levels are closely tied to specific symmetry sectors, resulting in strong correlations among energy levels within the same sector. However, there's typically little to no correlation between energy levels belonging to different symmetry sectors. The statistical analysis of nearest-neighbor level spacing reveals these correlations through the Wigner--Dyson distribution, while the spread complexity displays a peak just before reaching saturation. However, if we disregard the existence of different symmetry sectors within the system's Hamiltonian, even chaotic systems may appear to lack correlations between energy levels. Consequently, this oversight could result in the statistical distribution of nearest-neighbor level spacing resembling a mixture of the Wigner--Dyson and Poisson distributions, with the peak in spread complexity becoming less evident.
%


\bibliographystyle{JHEP}

\providecommand{\href}[2]{#2}\begingroup\raggedright\endgroup

\end{document}